\documentclass[11pt,a4paper]{article}
\usepackage{amsfonts,amssymb,graphics,epsfig,verbatim,bm,latexsym,amsmath,url,hyperref,amsbsy}
\usepackage{natbib}
\usepackage{tabto}
\usepackage{bbm}
\usepackage{xr}
\usepackage{booktabs}
\usepackage{subfig}
\usepackage{adjustbox}
\usepackage{multirow,multicol}
\usepackage{ragged2e}
\usepackage{lscape}
\usepackage{arydshln}

\usepackage{xcolor}
\definecolor{dgreen}{rgb}{0.,0.6,0.}
\definecolor{RSienna}{cmyk}{0,0.72,1,0.45}
\definecolor{dyellow}{rgb}{1,0.75,0}
\definecolor{dgreen}{rgb}{0.2,0.6,0.2}

\renewcommand{\thesection}{\arabic{section}}
\renewcommand{\theequation}{\arabic{section}.\arabic{equation}}

\begin{document}

\title{\normalfont\textbf{\fontsize{13}{14}{Shocking concerns: public perception about climate change and the macroeconomy}}}
\author{\hspace{-1.5cm} G. Angelini$^\dagger$, \hspace{0.05cm} M.E. Bontempi$^\dagger$, \hspace{0.05cm} L. De Angelis$^\dagger$, \hspace{0.05cm} P. Neri$^\dagger$, \hspace{0.05cm} M.M. Sorge$^\ddagger$ \\
\footnotesize{\hspace{-1.5cm} $^\dagger$University of Bologna, $^\ddagger$University of Salerno and CSEF} 
}
\date{\today}
\maketitle

\begin{abstract}
Public perceptions of climate change arguably contribute to shaping private adaptation and support for policy intervention. In this paper, we propose a novel Climate Concern Index (CCI), based on disaggregated web-search volumes related to climate change topics, to gauge the intensity and dynamic evolution of collective climate perceptions, and evaluate its impacts on the business cycle. Using data from the United States over the 2004–2024 span, we capture widespread shifts in perceived climate-related risks, particularly those consistent with the postcognitive interpretation of affective responses to extreme climate events. To assess the aggregate implications of evolving public concerns about the climate, we estimate a proxy-SVAR model and find that exogenous variation in the CCI entails a statistically significant drop in both employment and private consumption and a persistent surge in stock market volatility, while core inflation remains largely unaffected. These results suggest that, even in the absence of direct physical risks, heightened concerns for climate-related phenomena can trigger behavioral adaptation with nontrivial consequences for the macroeconomy, thereby demanding attention from institutional players in the macro-financial field. 
\\
\\
\textbf{Keywords:} Climate concerns; Identification; Macroeconomic outcomes; Financial stability; Proxy-SVAR
\\
\\
\textbf{JEL classification:} C5; C8; D83; E6; Q54
%\newline
\end{abstract}

\newpage

\section{Introduction}

In recent decades, the consequences of climate change have become increasingly visible, with extreme weather events, rising sea levels, and ecological disruptions. While diffuse scientific consensus has long emerged on the urgent need to mitigate climate change worldwide, public perceptions of climate variability remain crucial in shaping private adaptation and support for public sector intervention in both the environmental and economic landscapes. A growing body of research in fact suggests that public sentiment about current and future harms from climate change can influence not only environmental attitudes, but also consumption, saving, and investment choices and outcomes \citep{engle2020hedging,bernstein2019disaster,kruttli2021pricing}. On one hand, a broader understanding of climate issues may foster demand for sustainable products and services, promote the adoption of clean technologies, and accelerate the transition to a climate-resilient economy in the long term. On the other, in the absence of effective risk-sharing arrangements, rising anxiety or pessimism about climate-related risks and the impact on livelihood security and health may exacerbate precautionary behavior of economic players, potentially dampening aggregate economic activity in the shorter term. At the same time, financial market participants are increasingly incorporating climate-related risks and narratives into core portfolio and trading strategies \citep{hong2019climate}, while households have started intensifying their commitment to purchasing behavior that rewards environmentally responsible companies \citep{white2019shift}. 

The macroeconomic implications of collective climate concerns extend well beyond shifts in environmental preferences. Heightened levels of perceived climate risk may trigger additional uncertainty, which in turn affects expectations about future income, regulatory interventions, or even the long-term viability of entire industries. In this sense, climate concerns among the general public may operate similarly to uncertainty or sentiment shocks affecting precautionary saving, delaying consumption and investment decisions, and distorting educational and occupational choices. For instance, households facing growing concern about climate-related disruptions may increase savings and/or bequests as a buffer against current and future income uncertainty, or adopt more conservative spending patterns. Firms, likewise, may postpone capital expenditures or redirect resources toward sectors perceived as more resilient or aligned with the green transition. These behavioral adjustments, when aggregated, can generate measurable fluctuations in aggregate demand, labor utilization, and investment dynamics, thus influencing the business cycle. In addition, climate concerns can interact with policy preferences and expectations of policy strategies to adapt to or mitigate climate change. As public awareness increases, so may the perceived likelihood of new regulations, carbon pricing, or green subsidies, against which firms and households may adjust their current behavior in ways that produce immediate macroeconomic effects. Since optimal private adaptation is arguably hindered by several socio-economic frictions tied to, e.g., unequal access to investment opportunities and missing insurance markets, uncovering the key features of such forward-looking behavioral responses appears to be crucial for the design and implementation of targeted policy intervention \citep{carleton2024adaptation}. Despite these considerations, measures of climate perceptions remain scarce, and empirical studies quantifying their macroeconomic effects are limited. 

In this paper, we construct a novel Climate Concern Index (CCI), which captures the intensity and dynamics of general sentiment about and interest in climate change using disaggregated Google search data. Inspired by recent approaches that use online activity as a proxy for economic sentiment \citep{bontempi2021eurq,shields2023}, our index leverages search volumes for targeted climate-related queries that are likely to elicit concerns about, fears for and/or interest in climate-related issues, including references to extreme weather, natural disasters, and broader environmental risks. The CCI reflects bottom-up engagement with climate issues and offers high-frequency insights into how individuals perceive climate-related threats. While amenable to use within and across different geographical partitions, our analysis focuses on the United States over the 2004–2024 period, covering historical junctures and areas where substantial skepticism about the reality and the severity of climate change have arisen and grown, particularly across right-leaning constituencies. Within this context, particular attention is paid to the post-cognitive interpretation of affective responses \citep{vanderlinden2014personal}, whereby emotional concern arises after individuals attribute personal experiences to severe climate events - such as heatwaves, floods and hurricanes. This distinction reinforces the relevance of our index as a measure of endogenous sentiment shifts rather than exogenous physical events. We validate the CCI by comparing it with a range of alternative sentiment measures proposed in the literature, including news-based indices \citep{faccini2023,bua2024,engle2020hedging,ardia2023,gavriilidis2021measuring,noailly2022does} and social media-based metrics \citep{arteaga2024,santi2023}. 

We then assess the macroeconomic consequences of changes in climate concern using a proxy structural vector autoregression (SVAR) model \citep{stock2018identification,mertens2013dynamic,angelini2019exogenous}. The proxy-SVAR approach allows us to identify the causal impact of climate concern shocks on key macroeconomic variables without imposing overly restrictive or questionable identifying assumptions on the system. Robustly across a variety of alternative specifications of the baseline linear VAR, our findings indicate that exogenous increases in the CCI are associated with statistically significant declines in employment, private consumption spending and headline inflation, and thus appear to propagate throughout the economy as demand-type shocks. Core inflation, by contrast, remains largely unaffected, suggesting that price variation in food and energy items drives the bulk of the response of inflation to increased concerns about the climate. 

To gauge the effects of varying climate-related concerns on stock market volatility, we also explore how CCI shocks impact on the VIX index and find - in line with extant literature e.g. \cite{li2024study} - that expected market volatility slowly rises with widespread public awareness of climate change and then persist over a relatively long horizon. Given the major role played by stock markets in diversifying and alleviating financial risk, this finding lends support to the view that behavioral responses of financial market participants to climate variability qualify as a non-negligible threat to financial stability, and therefore demand attention from financial regulators and other institutional players in the macro-prudential field.

Overall, our empirical results suggest that collective perceptions of evolving climate change issues — absent any aggregate economic shock — can produce tangible macroeconomic effects through behavioral channels. By providing a new tool for tracking climate sentiment and demonstrating its relevance for macroeconomic outcomes, our paper thus contributes to the growing literature on the financial and economic implications of environmental risks. 

The remainder of the paper is structured as follows. Section \ref{sez:CCI} describes the construction of the Climate Concern Index. Section \ref{sez:comp} compares our index with alternative climate-related sentiment measures. Section \ref{sez:proxy_svar} presents the empirical methodology and results. Section \ref{CONCL} concludes.

\section{The Climate Concern Index}\label{sez:CCI}

The Climate Concern Index (CCI) is constructed by following the methodology in \citet{bontempi2021eurq} and relies on the volumes of web searches for an extensive list of disaggregated queries on climate change topics that may generate concern among economic agents. As forcefully argued in \cite{bontempi2021eurq}, internet search volumes can be thought of as reflecting people’s epistemic uncertainty - a \textit{lack of knowledge about potentially knowable facts} -- to mean that people's need for gathering enriched information emerges when they are concerned about something that is undetermined or unsettled and could directly or indirectly affect them.

The advantages of our approach are many. First, it is inherently meant to extract spontaneous and widespread public reactions, which are less influenced by survey biases or reporting delays vis-\`a-vis alternative measures based on survey data (e.g. the Climate Perception Index constructed by Meta in partnership with Yale).\footnote{The Climate Perceptions Index is based on annual data collected by Meta in its Facebook Climate Change Opinion Survey exploring three dimensions of individuals’ climate change knowledge: awareness, risk perception, and commitment to action. See: \scalebox{0.6}{\small \url{https://public.tableau.com/app/profile/petra1128/viz/CPIvisualizations\_update/ClimatePerceptionsIndex}}.} Second, in sharp contrast with indicators built on media and social networks, like newspapers and Twitter (see the comparison in section \ref{sez:comp}), our index entails a shift of focus from well-defined segments of the economy (e.g. financial market participants) to the average individual, insofar as it provides information on collective interests and concerns of millions of users that is both dynamic and granular. Third, it is based on freely accessible data updated in real time, and can in principle be computed for different geographical locations and political units all over the world\footnote{The data are even available at a minute frequency. Recently, \citet{Puhr2024} have used Google Trends as uniquely representative of grassroots sociopolitical sentiment, timely, and widely available.}

\subsection{The CCI vocabulary}

To gauge climate risk feelings and concerns among internet users, the definition and validation of search terms is of paramount importance. The construction of the Climate Concern Index (CCI) begin with a detailed and iterative process of vocabulary development. This step is crucial to ensure that the index accurately captures the diverse dimensions of climate-related concerns as expressed by the general public through web searches. We initially drew upon existing literature on climate sentiment indices to compile a preliminary list of relevant search terms. To refine and expand this list, we used prompt engineering techniques to interact with large language models, such as the one behind the ChatGPT chatbot by OpenAI. These models helped identify emerging and contextually relevant terminology used in climate discourse. In parallel, we examined content from specialized posts on social media and online platforms to validate the real-world salience of the proposed terms. Many candidate queries that appeared theoretically relevant turned out to be marginal in actual search frequency, while others, less expected, proved to be central to public concern. After multiple rounds of testing and validation, we selected a final set of 112 queries focused on the United States reported in Appendix \ref{sez:APPdic}. To better structure the index and allow thematic interpretation, we classified the selected queries into seven categories reflecting different affective or informational tones. These are: 
\begin{enumerate}
    \item natural disasters that generate fear (e.g., “extreme weather”, “floods”); 
    \item global warming-related threats (e.g., “rising seas”, “deforestation”);
    \item broader climate change narratives (e.g., “climate refugees”, “climate crisis”); 
    \item mitigation and reduction strategies (e.g., “climate solutions”, “circular economy”); 
    \item technological hope (e.g., “biodiversity”, “aquaculture”); 
    \item international summits (e.g., “Paris Agreement”, “Kyoto Protocol”); 
    \item climate or environmental policy terms (e.g., “decarbonization”, “IPCC”).
\end{enumerate}

\begin{figure}[!ht]
\centering
\emph{US fear \& concern} \\
{\includegraphics[scale=0.10]{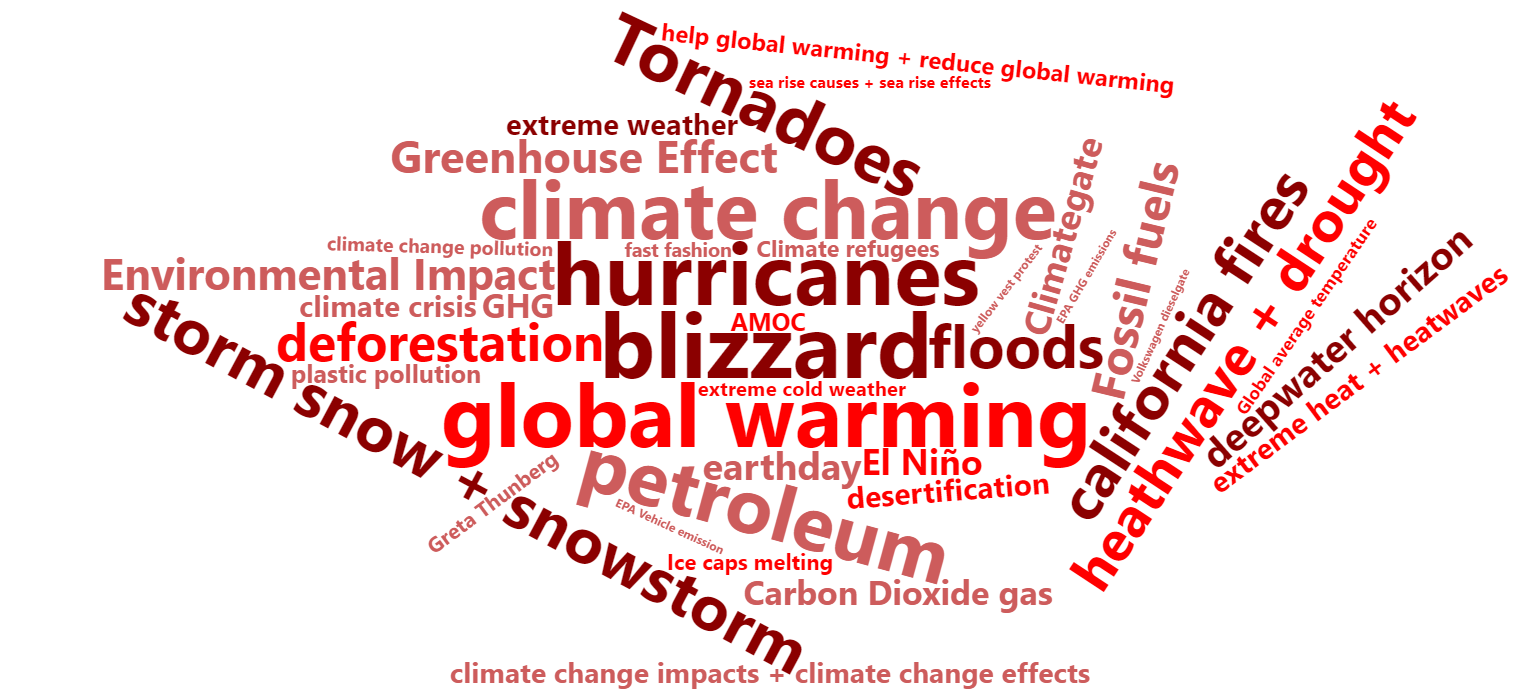}} \\
\emph{US policy \& summit}\\
{\includegraphics[scale=0.10]{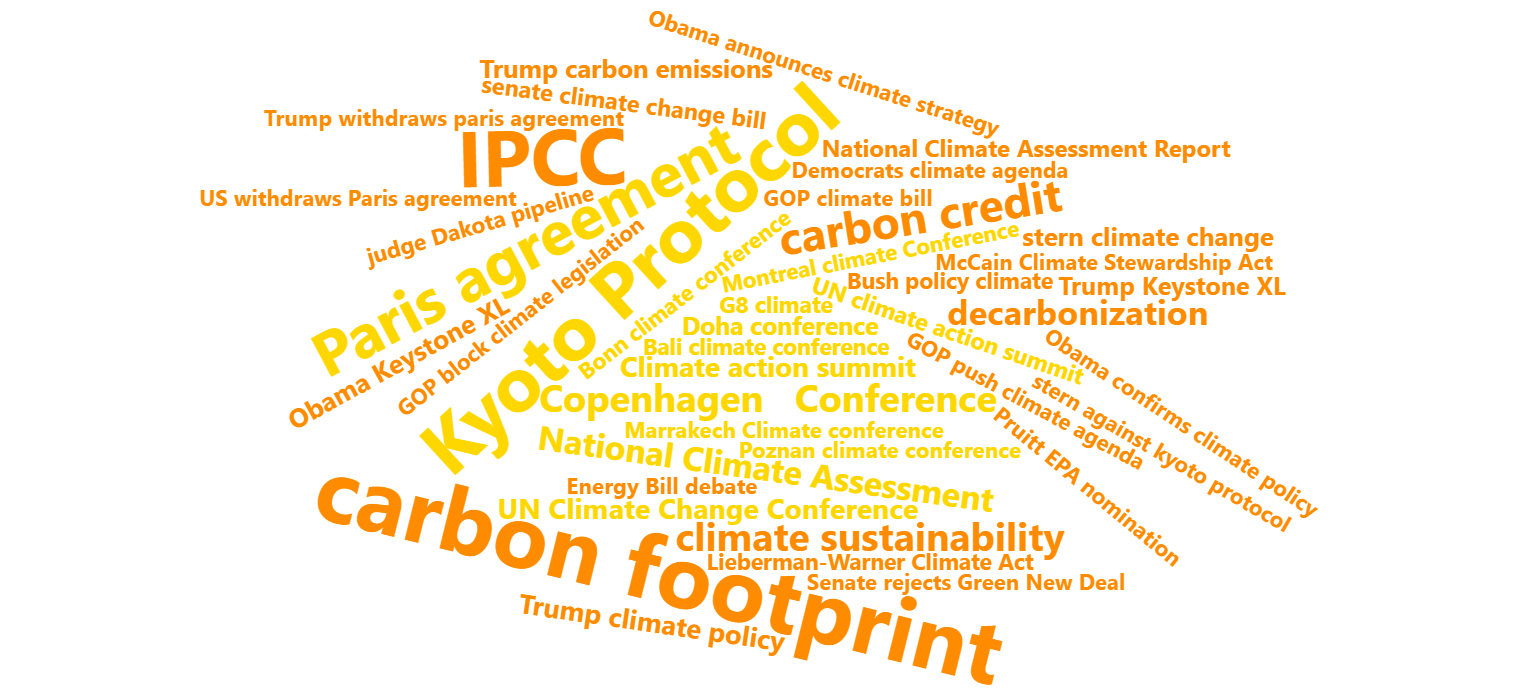}}  \\
\emph{US reduction \& hope}\\
{\includegraphics[scale=0.10]{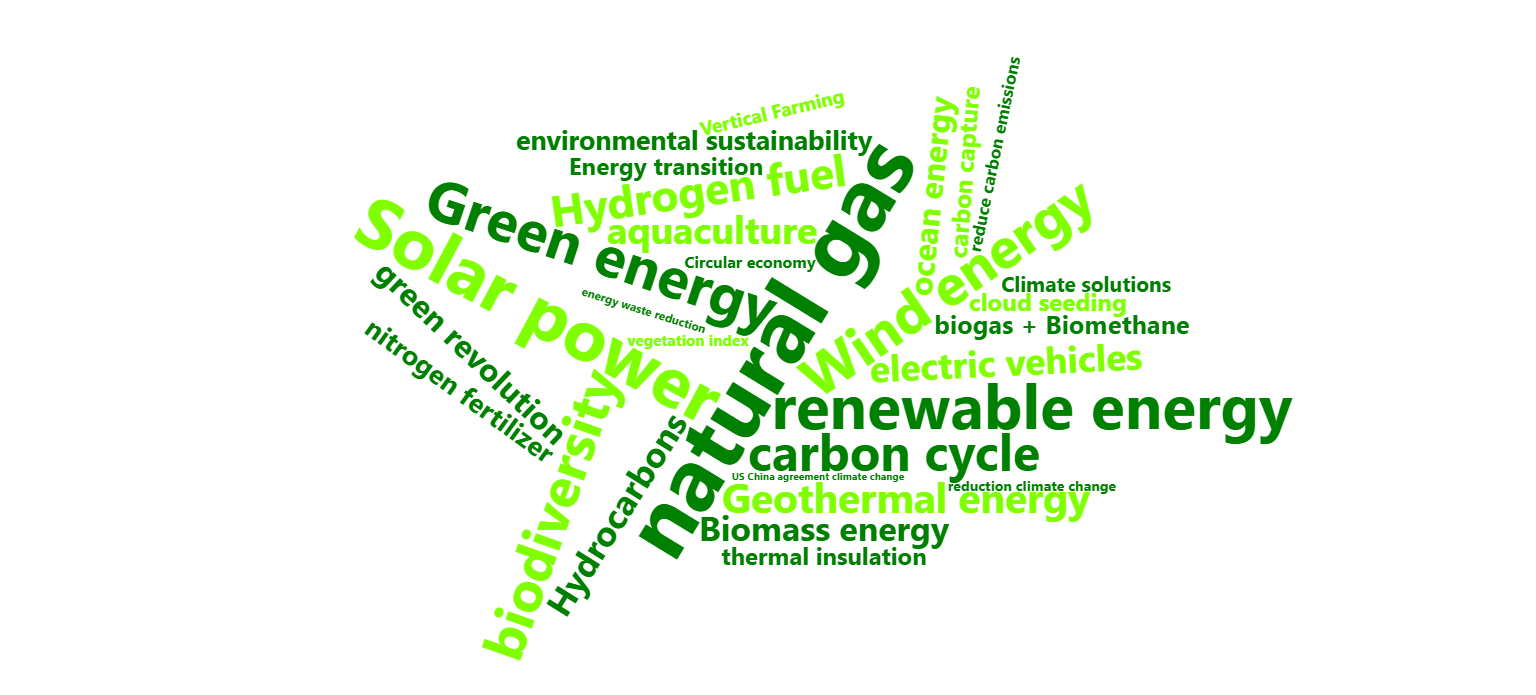}} 
\caption{Word clouds}
\label{fig:CLOUDS}
\end{figure}

The full list of queries is reported in Appendix~\ref{sez:APPdic}, along with their classification. This categorization allows us to explore not only the volume of people' informational queries for climate-related issues, but also its qualitative composition—distinguishing, for instance, between fear-based and solution-oriented attention. This distinction is particularly relevant for the economic interpretation of the mechanisms by which climate sentiment may influence macroeconomic outcomes. Figure~\ref{fig:CLOUDS} illustrates the most frequently searched terms across the seven thematic categories of the CCI, using word clouds as a visual representation. While word clouds are typically employed in textual analysis, here the size of each term reflects its relative frequency within the overall index. In 2024, for the United States, the most prominent category is climate change reduction and mitigation (category 4), accounting for 28\% of total searches. This is followed by natural disasters (category 1, 20\%), new technologies (category 5, 19\%), and general climate change concerns (category 3, 19\%). Global warming-related terms (category 2) contribute 10\%, while international summits (category 6) and environmental policies (category 7) jointly account for the remaining 4\%.

Extreme weather events are among the most frequently searched topics in our CCI, highlighting the central role that direct experiences of climate-related phenomena play in shaping public concern. This is consistent with the post-cognitive interpretation proposed by \citet{vanderlinden2014personal}, according to which individuals must first cognitively associate an event, such as a hurricane, flood, or heatwave, with climate change in order to experience negative affective reactions. These reactions are strong determinants of perceived climate risk and are a key driver of information-seeking behavior \citep{campiglio2025climate}. More broadly, public engagement with climate change is often hindered by its perceived psychological distance. As noted by \citet{vanderlinden2015}, many individuals view climate change as a temporally, spatially, and socially distant threat, which can lead to procrastination in both personal and collective decision-making regarding mitigation and adaptation strategies. This gap between scientific urgency and perceived immediacy underscores the importance of tools like the CCI, which help quantify how concern evolves over time and in response to salient events. The relatively limited share of search activity dedicated to international summits (category 6) and environmental policy terms (category 7) may reflect the diffuse and often fragmented nature of climate governance. While many countries aim to formally adopt net-zero targets and long-term strategies to limit global warming, the transition paths remain highly uncertain. In particular, the Network for Greening the Financial System (NGFS) outlines several climate scenarios based on varying degrees of physical and transition risks. Among them, disorderly transition scenarios, characterized by delayed or inconsistent policy action, can generate significant economic dislocations and increase the exposure of financial and economic systems to climate-related shocks.\footnote{See NGFS Scenarios Portal: \url{https://www.ngfs.net/ngfs-scenarios-portal/}} Given the absence of a globally coordinated climate mitigation framework, future policy trajectories are inherently uncertain. This uncertainty can affect public expectations and potentially influence both individual and firm behavior, which further justifies the need for high-frequency, sentiment-based indicators like the CCI.

\subsection{CCI construction}

The first step in constructing the Climate Concern Index (CCI) is the selection of a suitable benchmark query. The choice of such a benchmark is crucial in any Google Trends-based index, as it governs the rescaling of the search intensities across different groups. Following \citet{castelnuovo2017google}, a good benchmark should be frequent enough to ensure comparability across time, but not too volatile or closely tied to the specific events the index is intended to capture. We tested alternative candidates, starting with ``global warming'' (the only query used by \cite{choi2020}) and ``climate change'' (the only query used by \cite{ding2022}). One incontrovertible fact that emerges is how the Paris Conference in December 2015 shifted general interest from “global warming” to “climate change”. For example, in the US, starting from 2016:M1, ``global warming'' decreased from an average 13.37\% to 4.72\%, while ``climate change'' increased from 4.46\% to 10.4\% in their relative weight within the CCI. Consequently, we opted for a more neutral and stable benchmark: ``natural gas''. This term maintains a relatively constant search frequency over time (averaging around 13.4\%) and, while environmentally relevant, is less directly tied to event-driven surges in concern. 

\begin{figure}[!ht]
\centering
\includegraphics[scale=0.25]{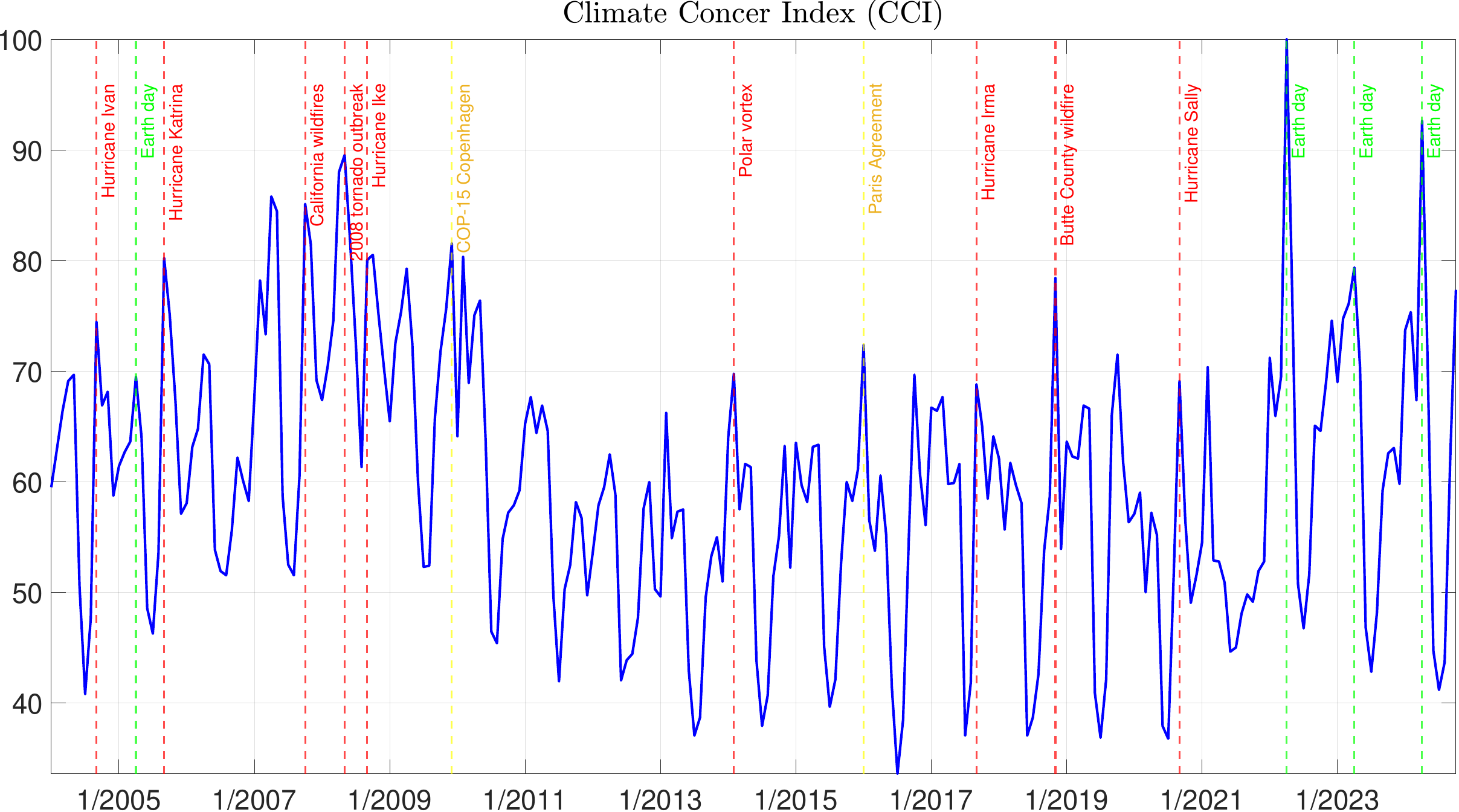}
\caption{\small{Climate Concern Index (CCI) in \eqref{eq:cci}.}} 
\label{fig:USIT1}
\end{figure}

Having selected the benchmark, we proceeded to organize the remaining search queries into a structured set suitable for index construction. Specifically, we grouped the queries $i = 1, \dots, I$ into subsets $j = 1, \dots, J$, ordered and homogeneous according to the seven thematic categories. Each group consists of a maximum of five terms and must contain the benchmark. We defined the time interval $[t_1, t_2]$ for $t_1$ = 2004:M1 to $t_2$ = 2024:M10, and the country of interest $C = \text{US}$. We then computed the relative frequency index $FI$ for each term $i$ within group $j$ as follows:
\[
FI_{[t_1,t_2]}^{C}(i \in j) = 100 \times \frac{S_{[t_1,t_2]}^{C}(i \in j)}{\underset{[t_1,t_2], j}{\max}\left (S_{[t_1,t_2]}^{C}(i)\right)} \qquad \forall \ j \in J, \quad \forall \ i \in I
\]
where $S_{[t_1,t_2]}^{C}(i \in j)$ is the time series of the relative search volume for term $i$ within group $j$, and $\underset{[t_1,t_2], j}{\max}\left(S_{[t_1,t_2]}^{C}(i)\right)$ is the search volume of the benchmark query over the same period. Each time series is bounded between 0 and 100. Once all relative frequency indices $FI^C_{[t_1,t_2]}$ are computed for each query within their respective groups, we aggregate them to construct the overall index, as follows:
\begin{equation}
CCI^C_{[t_1,t_2]}= \sum_{i=1}^{I} FI_{[t_1,t_2]}^{C}(i). \label{eq:cci}
\end{equation}
The resulting series is then normalized so that its maximum value over the sample period equals 100. This aggregation yields a time series of climate concern that reflects the dynamic intensity of search activity across the selected vocabulary, normalized and benchmarked by construction. As these are relative search volumes with respect to the total number of Google searches in a given period $[t_1,t_2]$, they are less sensitive to fluctuations in overall internet usage. However, in some rounds it may happen that the selected benchmark is not the maximum (e.g., a highly salient term like ``blizzard'' may spike to 100 while the benchmark is only at 50). From Figure \ref{fig:USIT1} we observe clear peaks corresponding to extreme weather events. A hurricane, for example, may trigger the cognitive link between the event and climate change, generating affective responses that drive search activity.\footnote{This mechanism appears to reflect the post-cognitive view put forward in \citet{vanderlinden2014personal}, whereby people develop concern only after attributing personal experiences to broader climate phenomena.}

From Figure \ref{fig:USIT1}, one can distinguish three distinct phases: an initial growth in climate concern from 2004 to 2010, a stagnation between 2011 and 2020, possibly linked to rising skepticism, and a renewed upward trend beginning in late 2021. According to \citet{Baiardi2023}, this latter phase may be partially driven by climate communication through social media, amplified by popular figures such as Mr. Beast or Greta Thunberg. “Earth Day” events, held annually on April 22 since 1970, also tend to produce noticeable spikes in the index. These spikes vary in intensity depending on the year's theme and media resonance, and they offer a way to detect the influence of coordinated environmental campaigns on public sentiment \citep{Leombruni2015}.\footnote{See: \scalebox{0.6}{\small\url{https://www.earthday.org/}} and \\ \scalebox{0.6}{\small\url{https://growbilliontrees.com/blogs/knowledge/earth-day-themes-over-the-years-key-messages-and-movements-for-a-greener-planet}}.}

\section{A comparison with other indices}
\label{sez:comp}

In this section we compare our CCI with the other climate-related indices proposed by the literature mainly for the US and used to evaluate risk premia in financial markets related to climate change. We can distinguish three main groups: (1) the Reuters-based measures of \citet{faccini2023,bua2024}; we select \citet{faccini2023} as it is updated and matches the period covered by CCI for the US. (2) The newspaper-based measures of \citet{engle2020hedging,ardia2023,gavriilidis2021measuring,noailly2022does}; we select \citet{ardia2023,gavriilidis2021measuring} as they cover most of the sample period of the CCI for the US, while \citet{engle2020hedging,noailly2022does} are not regularly updated. (3) The Twitter-based measures of \citet{arteaga2024,santi2023}, which cover reduced samples, beginning in 2014 and 2010 respectively, with \citet{santi2023} not regularly updated. A detailed overview of these indices is reported in Table \ref{tab:otherindices}, showing that most are based on newspapers. The indices most thoroughly analyzed in comparison with our CCI are highlighted in bold.\footnote{It is worth mentioning two papers that rely on web searches but use only a single, partial query. \citet{ding2022} measure investor attention through the Google Search Volume Index (GSVI) of "climate change" and find significant causal effects on spillovers among carbon, fossil energy, and clean energy markets. \citet{choi2020} use the GSVI of "global warming" as a proxy for investor reaction, showing that warmer-than-usual temperatures lead individuals to revise their beliefs about climate change. In financial markets, stocks of carbon-intensive firms underperform low-emission firms during abnormally warm periods.}

\begin{table}[!htp] 
\caption{Other climate risk indices}\label{tab:otherindices} 
\centering
\tiny
\begin{tabular}{llllll}
\toprule
Paper                 &Country&Period&Source  \\
\midrule
\textbf{\citet{faccini2023}} &US&2000:M1-2023:M6&Reuters climate-change news \\
\citet{bua2024}&Europe&2005:M1-2022:M12&Reuters climate-change news \\
\citet{engle2020hedging}&US&1984:M1-2017:M6&Wall Street Journal\\
\textbf{\citet{ardia2023}}&US&2003:M1-2024:M6& 10 newspapers\\
\textbf{\citet{gavriilidis2021measuring}}&US&1987:M4-2024:M6& 8 newspapers\\
\citet{noailly2022does}&US &1981:M1-2019:M3& 10 newspapers \\
\textbf{\citet{arteaga2024}}& 8 countries&2014:M10-2022:M12&Twitter\\
\citet{santi2023}& US&2014:M10-2019:M9& StockTwits posts\\
\bottomrule
 \end{tabular}
 \newline
\newline
\end{table}

\citet{faccini2023} exploit Latent Dirichlet Allocation-based textual analysis to quantify the intensity of transition risks elicited by the U.S. political debate related to natural disasters, global warming, international summits, and climate policy; only the climate-policy factor is priced, especially post-2012, and the risk premium is consistent with investors hedging the imminent transition risks from government intervention, rather than the direct risks from climate change itself. \citet{bua2024} present transition risk (technological advances and environmental policies) and physical risk (extreme and chronical hazards directly caused by climate change); economically significant transition risk premium emerges post-2015 in Europe. \citet{engle2020hedging} propose a textual analysis-based algorithm that maps news articles into two monthly time series that proxy the latent shocks in the attention and negative attention about climate change; a mimicking portfolio approach can be successful in hedging innovations in climate change news. In \citet{ardia2023}, on days with an unexpected increase in climate change concerns, the green firms’ stock prices tend to increase while brown firms’ prices decrease; using topic modeling, this effect holds for concerns about both transition and physical climate change risk. More related to climate policy uncertainty than climate change concern in general are \citet{gavriilidis2021measuring} that has a strong and negative effect on CO2 emissions; and the three indices of sentiment on environmental policy, renewable energy policy and international climate negotiations of \citet{noailly2022does}: environmental policy is associated with a greater probability of clean technology startups receiving venture capital (VC) funding and reduced stock returns for high-emissions firms most exposed to environmental regulations. In \citet{arteaga2024} a country experiencing more severe climate news shocks tends to see both an inflow of capital and an appreciation of its currency; these outcomes align with the expectations of a risk-sharing model in which investors price climate news shocks and engage in the trade of both consumption and investment goods. In \citet{santi2023} stocks of emission (carbon-intensive) firms underperform clean (low-emission) stocks when investor climate sentiment is more positive, with investor overreaction to climate change risk and reversal in longer horizons; salient but uninformative climate change events, such as the release of a report on climate change and abnormal weather events, facilitate the investor learning process and correction of the mispricing. Albeit the construction of most of these indexes share similar rationale and approaches, the correlation between these indexes in general does not exceed 0.5 (see, e.g., Table 1 in \citet{ardia2023}).
\citet{faccini2023} present four indices focusing on different aspects of climate change: climate policy, global warming, international summits and natural disasters. 

%Among them, seasonality is significantly present only in international summits (obviously given the periodic nature of these events), hence we seasonally adjusted this series with Demetra. Despite persistence (first-order autocorrelations are in the range 0.5-0.8), all series are at least 10\% stationary in the light of both ADF tests (with SBC lag selection), and Johansen cointegration rank equal to four in a VAR(2), again with SBC selection. The correlation coefficients between the four indices suggest that international summits are the only variable weakly ($<31\%$) correlated with the others, while climate policy, global warming, and natural disasters can be condensed by the first principal component explaining about 74\% of their total variance. 

\begin{figure}[ht!]
\centering
\includegraphics[scale=0.50]{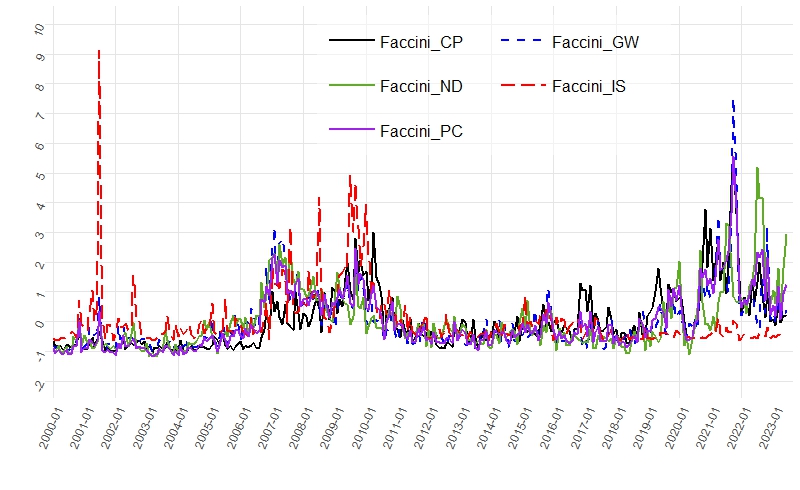} 
\caption{Standardized \citet{faccini2023} indices (IS, international summits seasonally adjusted, PC first principal component of climate policy, CP, global warming, GW, and natural disasters, ND).}\label{fig:reu}
\end{figure}

Figure \ref{fig:reu} shows the five standardized indices: Faccini\_IS (seasonally adjusted) and the first principal component, Faccini\_PC, extracted from Faccini\_GW, Faccini\_ND and Faccini\_CP: we observe an increase in levels over the period 2018-2023, excluding the first half of the Covid period.

\begin{figure}[ht!]
\centering
\includegraphics[scale=0.5]{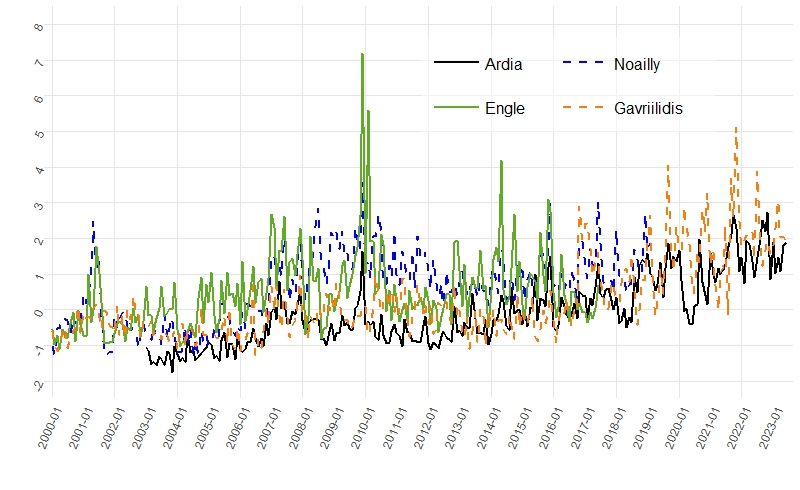} 
\caption{Standardized newspapers indices of \citet{engle2020hedging,ardia2023,gavriilidis2021measuring,noailly2022does}.}\label{fig:news}
\end{figure}

Figure \ref{fig:news} shows all available standardized newspapers-based measures \citep{engle2020hedging,ardia2023,gavriilidis2021measuring,noailly2022does} and suggests a common pattern in the overlapping sample period. Indeed, the correlation matrix between the various series shows relationships on the order of 0.6-0.7. %All measures exhibit no seasonality and are stationary in the light of both the ADF unit root test and the Johansen test of full rank (unlike the Reuters case, a deterministic trend is added to account for the growing pattern of concern over time).
Principal components’ analysis of \citet{ardia2023,gavriilidis2021measuring} delivers that the first principal component explains about the 86\% of the total variance of the two series. Remarkably, despite the fact that \citet{gavriilidis2021measuring} and \citet{ardia2023} are measures of uncertainty and climate change, respectively, they nonetheless share relevant common features, with a correlation coefficient of about 0.71.

\begin{figure}[ht!]
\centering
\includegraphics[scale=0.5]{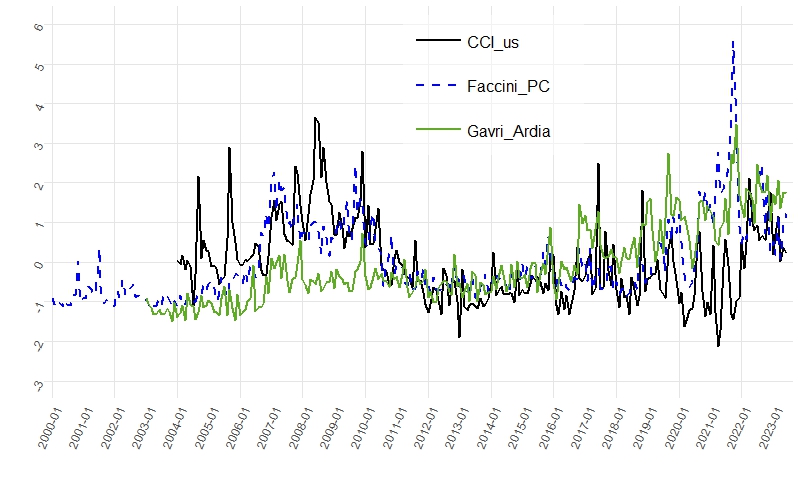} 
\caption{Standardized indices from Reuters (first principal component of \cite{faccini2023} climate policy, global warming and natural disasters), newspapers (first principal component of \cite{ardia2023,gavriilidis2021measuring}) and CCI for the US.}\label{fig:reunewsCCI}
\end{figure}

Therefore, Figure \ref{fig:reunewsCCI} compares faccini\_PC, the first principal component of the Reuters-based indices (climate policy, global warming and natural disasters of \cite{faccini2023}) and Gavri\_Ardia, the first principal component of the newspapers-based indices \citep{ardia2023,gavriilidis2021measuring}, and suggests a common increase after 2018, which stops around the end of 2021. In contrast, the rise that occurred in the Reuters-based measure around 2006-2007 was not accompanied by a similar increase in the newspapers-based measure. The Figure also reports our seasonally adjusted standardized CCI for the US, as these three series are available for a common period that allows us to assess the leading/lagging properties of the indices through Granger causality.\footnote{Results are robust to the inclusion of \citet{faccini2023} index focused on international summits.}

It is interesting to note that many ``journalistic'' terms used by other authors to construct their indices are less visible than other queries in the CCI. For example, references to IPCC reports \citep{faccini2023} produced only small spikes (e.g., 2007:M2, 2018:M10, and April 2014, 2021--2023). The EPA administrator Scott Pruitt generated modest peaks in 2016:M12 and 2018:M4, likely due to his controversial stance. The ``Keystone pipeline bill" query peaked in 2014:M11 and 2015:M2, with only a minor increase in 2017M1, despite its political relevance. \citet{vanderlinden2017} notes that many individuals struggle to understand the mechanisms of climate change, misidentify causes and effective solutions, and may conflate it with broader environmental issues. The link between knowledge and concern appears to be moderated by political ideology \citep{Shao2016,McCright2011}. The perception of climate risks also depends on experience, particularly with extreme weather (e.g., hurricanes, floods, heatwaves), which is more tangible than gradual temperature rise \citep{Pawlik1991}.

We exploit Granger causality to assess the leading/lagging properties of the three indices in Figure \ref{fig:reunewsCCI}. Given a time span of about 240 months, we estimated a VAR(13) model which, although inefficient, measures all monthly dynamics within one year. As no root is found outside the unit circle, the VAR satisfies the stability condition and residuals autocorrelation tests were never significant up to the $13^{th}$ order. 
In this VAR representation, the in-sample predictability of Reuters and newspapers indices is much better than that of the CCI, with $R^2$ being 0.79 and 0.83 respectively, compared to 0.42 for the latter. This suggests that random shocks to CCI explain more that half of its variability, while for the other two indices variability can be better explained by past shocks to all variables.
The simultaneous correlation of the different shocks shows that, within the month, the Reuters and newspapers shocks are fairly correlated (around 0.4), while the occurrence of the CCI shocks is not highly correlated with either Reuters or the newspapers (0.12 and 0.25, respectively). This suggests that the CCI time series incorporates some information not measured by the other two variables.
Overall, out Granger causality tests, reported in Table \ref{tab:VAR}, indicate that newspapers’ attention about climate change and uncertainty about climate policy lead Reuters climate change news and vice versa (i.e. Reuters leads newspapers): a sort of bilateral Granger causation. Instead, and most importantly, the CCI (people interest through Google searches) is not significantly caused (led) by past information on the newspapers and Reuters. 

\begin{table}[!ht] 
\caption{Granger causality in the US VAR}\label{tab:VAR}
\centering
%\begin{adjustbox}{scale=0.9}
\begin{adjustbox}{max width=\textwidth}
    \begin{tabular}{llll}
\toprule
& Chi-sq & df & Prob. \\
\hline
\multicolumn{4}{l}{Dependent variable: Principal component of \citet{faccini2023}} \\ 
\hline
CCI for the US  & 18.015 & 13 & 0.1570 \\
Principal component of \citet{ardia2023,gavriilidis2021measuring}  & 26.356 & 13 & 0.0152 \\
All & 49.146	&26	&0.0040 \\
\hline
\multicolumn{4}{l}{Dependent variable: CCI for the US} \\
\hline 
Principal component of \citet{faccini2023}  & 22.407 & 13 & 0.0494 \\
Principal component of \citet{ardia2023,gavriilidis2021measuring}  & 15.941 & 13 & 0.2624 \\
All & 32.519	&26	&0.1765 \\
\hline
\multicolumn{4}{l}{Dependent variable: Principal component of \citet{ardia2023,gavriilidis2021measuring} } \\ 
\hline 
Principal component of \citet{faccini2023}  & 24.200 & 13 & 0.0293 \\
CCI for the US  & 14.048 & 13 & 0.3705 \\
All & 49.740	&26	&0.0034 \\
\bottomrule
\end{tabular}
\end{adjustbox}
\end{table}

Twitter-based indices \citep{arteaga2024,santi2023} offer interesting points of comparison and further support the idea that digital traces, whether from social media or search data, capture complementary aspects of climate concern dynamics. Unfortunately, they could not be included in the VAR due to shorter sample periods: \citet{arteaga2024} spans 2014:M10--2022:M12 and \citet{santi2023} covers 2010:M1--2019:M9. Still, the rise in all indices after 2018 remains a stylized fact.\footnote{Notably, \citet{arteaga2024}'s index begins to rise after the first half of 2020 and levels off by late 2021, coinciding with the onset of the Ukraine-Russia war and the associated energy crisis.}

\begin{figure}[ht!]
\centering
\includegraphics[scale=0.50]{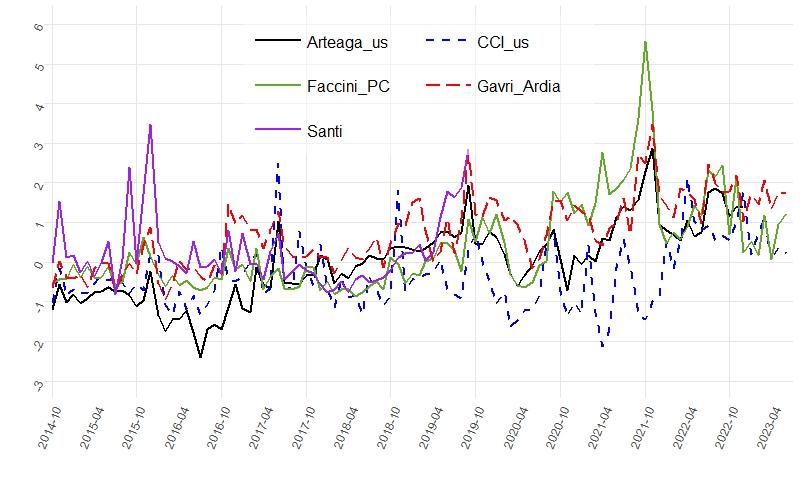} 
\caption{Standardized indices from Reuters (first principal component of \citet{faccini2023} climate policy, global warming and natural disasters), newspapers (first principal component of \cite{ardia2023,gavriilidis2021measuring}), Twitter \citep{arteaga2024,santi2023} and CCI for the US.}\label{fig:twinewsreu}
\end{figure}  

\section{Climate concern shocks and the macroeconomy}\label{sez:proxy_svar}
\subsection{Framework of analysis}

We now turn to estimating the macroeconomic effects of climate awareness feeding a structural VAR model with our synthetic measure of climate concerns along with key macroeconomic time series available at the monthly frequency. Our reference framework is the Proxy-SVAR model described in \cite{angelini2019exogenous}. Formally, we consider the following SVAR system:
\begin{equation}
Y_{t} = \Phi X_{t} + \eta_{t}, \quad \eta_{t} = B \varepsilon_{t}, \quad t = 1, \ldots, T, \label{eqvarreduced}
\end{equation}
where $Y_t$ is the $n \times 1$ vector of endogenous variables, $X_t = (Y_{t-1}^\prime, \ldots, Y_{t-l}^\prime)^\prime $ is the vector of $l$ lags of the variables, and $ T $ is the number of time observations. The matrix $ \Phi = (\Phi_1, \ldots, \Phi_k) $ is an $ n \times nk $ matrix containing the autoregressive coefficients, and $ \eta_t $ is an $n \times 1 $ vector of reduced-form VAR innovations. In this formulation, the system of equations $\eta_t = B \varepsilon_t $ maps the $n \times 1$ vector of structural shocks $\varepsilon_t$ to the reduced-form innovations via the $n \times n$ nonsingular matrix $B$, which contains the instantaneous (on-impact) coefficients. It is assumed that the structural shocks have a normalized covariance matrix $\Sigma_{\varepsilon} = \mathbb{E}(\varepsilon_t \varepsilon_t^\prime) = I_n $, meaning that we consider responses to one-standard-deviation shocks unless otherwise specified. Let $\varepsilon_{1,t}$ be the $k \times 1$ subvector of structural shocks $\varepsilon_t$, where $ 1 \leq k < n$, containing the target shocks of interest. The objective of the analysis is to identify and estimate the $h $-period-ahead responses of the variables $Y_{t+h}$ to the $j $-th target shock in $\varepsilon_{1,t} $:
\begin{equation}
IRF_{\bullet j}(h) = \frac{\partial Y_{t+h}}{\partial \varepsilon_{1,j,t}} = (S_n({C}_y)^h S_n^\prime)B_{\bullet 1}, \quad 1 \leq j \leq k, \label{eq_IRF}
\end{equation}
where $S_n$ is a selection matrix such that $S_n S_n^\prime = I_n$, ${C}_y$ is the companion matrix of the SVAR in \eqref{eqvarreduced}, and $B_{\bullet 1}$ is the first column of the on-impact matrix $B$. 

Given our goal of estimating the responses of the endogenous variables to the structural shock forcing the CCI dynamics, we place the CCI variable first in the vector \( Y_t \) and focus on identifying the first column of the matrix \( B \), called \( B_{\bullet 1} \). This task is performed by appealing to an external instrument \( z_t \), modeled as:
\begin{equation}
z_t = \phi \varepsilon_{1t} + \omega_t, \label{eq_instrument}
\end{equation}
where \( \phi \) is the relevance parameter, and \( \omega_t \) is a measurement error term uncorrelated with all the structural shocks. The partial shocks strategy described in \cite{angelini2019exogenous} allows identification of the first column of the structural matrix without imposing further restrictions on the other columns. Given the following relationships between the reduced-form and structural parameters:
\begin{eqnarray}
\Sigma_{z,\eta} \Sigma_{\eta}^{-1} \Sigma_{\eta,z} &=& \phi^2 \label{eq_moments_1} \\
\Sigma_{z, \eta} &=& \phi B_{\bullet 1}^\prime, \label{eq_moments_2}
\end{eqnarray}
where \( \Sigma_{z,\eta} \) is the covariance between the external instrument and the reduced-form residuals, and \( \Sigma_{\eta} \) is the covariance matrix of the reduced-form residuals, the estimation of these moments is performed using the Classical Minimum Distance (CMD) estimator by solving:
\begin{equation}
\min_{\theta} \left[ \sigma - f(\theta) \right]^\prime \Omega_{\sigma}^{-1} \left[ \sigma - f(\theta) \right], \label{eq_cmd}
\end{equation}
where \( \sigma =(\Sigma_{z,\eta} \Sigma_{\eta}^{-1} \Sigma_{\eta,z},\Sigma _{z,\eta})^{\prime } \) is the vector of reduced-form parameters, \( \theta =(\phi,B_{\bullet 1}^{\prime} )^{\prime} \) is the vector of structural parameters, \( f(\theta) \) maps structural parameters to reduced-form moments as in \eqref{eq_moments_1}-\eqref{eq_moments_2}, and \( \Omega_{\sigma} \) is the covariance matrix of \( \sigma \).
\\

\noindent{\textbf{Data and model specification}}

We feed the SVAR model (\ref{eqvarreduced}) with our CCI series along with a set of official statistics for the US economy, all available from the Federal Reserve Bank of St. Louis' FRED at the monthly frequency: the percentage change (and alternatively, the year-on-year growth rate) in industrial production, consumer price index (CPI) inflation, core CPI (CCPI) inflation, the short-term interest rate (the effective federal funds rate), the unemployment rate, the percentage change (and alternatively, the year-on-year growth rate) in Personal Consumption Expenditures (PCE), and the CBOE Volatility Index (VIX). We use data from 2004:M1 to 2023:M6 and all series have been seasonally adjusted.\footnote{We use six lags in the SVAR system, to gauge the persistence of the effects of exogenous changes in the CCI over time.}

We consider alternative specifications for the vector of endogenous variables $Y_t$ in the SVAR model \ref{eqvarreduced}, in order to estimate the responses of different components of total GDP (industrial productions vs. private consumption spending) or different indicators of consumer price inflation (including or excluding price variation in energy and food items) to exogenous variation in the CCI series. Specifically, we adopt the SVAR model for $Y_t = (cci_t, y_{2,t}, y_{3,t}, r_t, u_t)^\prime$, where $cci_t$ is our measure of climate concern, $y_{2,t} \in \left\lbrace ip_t, c_t \right\rbrace $ is either industrial production growth ($ip_t$) or aggregate consumption growth ($c_t$), $y_{3,t} \in \left\lbrace \pi_t, \pi^c_t  \right\rbrace$ is the inflation rate ($\pi_t$) or the core inflation rate ($\pi^c_t$), $r_t $ is the short-term interest rate, and $ u_t $ is the unemployment rate. To inspect the role of varying climate-related concerns on driving stock market volatility, we finally estimate a VAR model including the VIX index $v_t$, along with measures of real economic activity and price variation, i.e. we consider the alternative set of endogenous variables $Y_t = (cci_t, c_t, \pi^c_t, r_t, u_t, v_t)$.  

%The reduced-form residuals $ \eta_t $ and structural shocks $\varepsilon_t$ are linked by:
%\begin{equation}
%\begin{bmatrix}
%\eta_t^{cci} \\
%\eta_t^{y_2} \\
%\eta_t^{y_3} \\
%\eta_t^{r} \\
%\eta_t^{u} \\
%\eta_t^{v} \\
%\end{bmatrix} =
%\begin{bmatrix}
%b_{cci,cci} & b_{cci,y_2} & b_{cci,y_3} & b_{cci,r} & b_{cci,u}  & b_{cci,v}\\
%b_{y_2,cci} & b_{y_2,y_2} & b_{y_2,y_3} & b_{y_2,r} & b_{y_2,u} & b_{y_2,v}\\
%b_{y_3,cci} & b_{y_3,y_2} & b_{y_3,y_3} & b_{y_3,r} & b_{y_3,u} & b_{y_3,v}\\
%b_{r,cci} & b_{r,y_2} & b_{r,y_3} & b_{r,r} & b_{r,u} & b_{r,v}\\
%b_{u,cci} & b_{u,y_2} & b_{u,y_3} & b_{u,r} & b_{u,u} & b_{u,v} \\
%b_{v,cci} & b_{v,y_2} & b_{v,y_3} & b_{v,r} & b_{v,u} & b_{v,v} \\
%\end{bmatrix}
%\begin{bmatrix}
%\varepsilon_t^{cci} \\
%\varepsilon_t^{y_2} \\
%\varepsilon_t^{y_3} \\
%\varepsilon_t^{r} \\
%\varepsilon_t^{u} \\
%\varepsilon_t^{v}
%\end{bmatrix}, \label{eq_structural_system}
%\end{equation}
%where \( \eta_t \) is the vector of reduced-form residuals, and \( \varepsilon_t \) represents the structural shocks. The main goal is to identify the climate concern shock \( \varepsilon_t^{cci} \) and evaluate its dynamic effects on aggregate variables contained in $Y_t$ on the 2004:M1-2023:M11 time span. 

Given the mapping $\eta_t=B \varepsilon_t$ between structural shocks and reduced form innovations, our goal is to identify the CCI shock $\varepsilon_t^{cci} \in \varepsilon_t $ by instrumental information and then evaluate its dynamic effects on all the endogenous variables contained in $Y_t$ for the sample under scrutiny. To this end, we consider as instrument for $\varepsilon_t^{cci}$, i.e. the variable $z_t$ in \eqref{eq_instrument}, the change in the frequency of temperatures above the 90th percentile relative to the reference period (1961 to 1990). This variable, typically labeled as T90, is part of the ACI index developed by actuary associations in the United States and Canada, and used in \cite{kim2025severe} to uncover time-varying effects of severe weather variability on macroeconomic outcomes. Extreme temperature variation, in fact, is arguably exogenous with respect to the business cycle variables employed in the VAR and measured at the monthly frequency \citep{lucidi2024effects}. Remarkably, the chosen instrument proves to be strong (relevant) across all the employed specifications, according to the test proposed in \citet{angelini2024identification}.\footnote{The construction of the nation-wide T90 is based on raw gridded data, and requires (i) obtaining standardized temperature anomalies for any given location, computed as the difference between the $t$-dated observation and the average temperature for the same month during the reference period of 1961-1990, scaled by its reference period standard deviation); (ii) calculating percentiles based on the distribution of the standardized anomaly temperature observed at each grid point over the reference period; (iii) aggregating to the national level by averaging across all the stations on the grid within the US territory.}

\subsection{Results}

\begin{figure}[ht!]
\begin{center}
\begin{tabular}{cc}
\toprule
{\includegraphics[width=5cm,keepaspectratio]{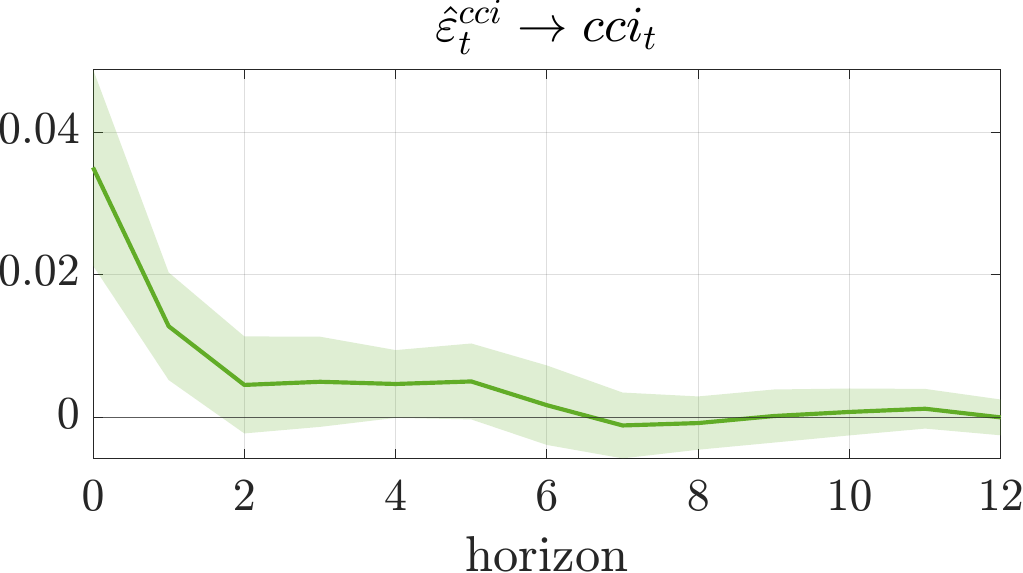}} & {\includegraphics[width=5cm,keepaspectratio]{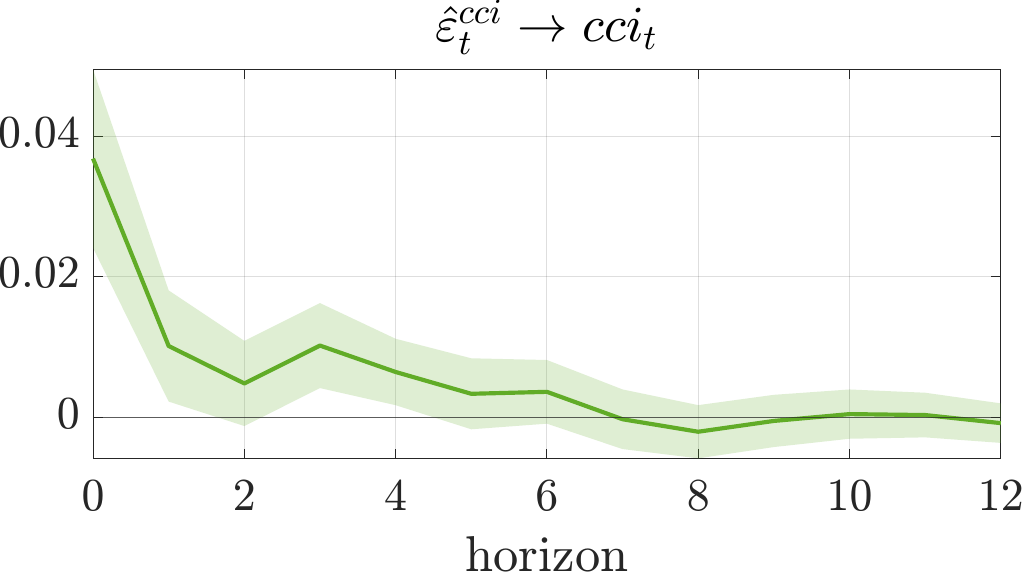}}   \\
{\includegraphics[width=5cm,keepaspectratio]{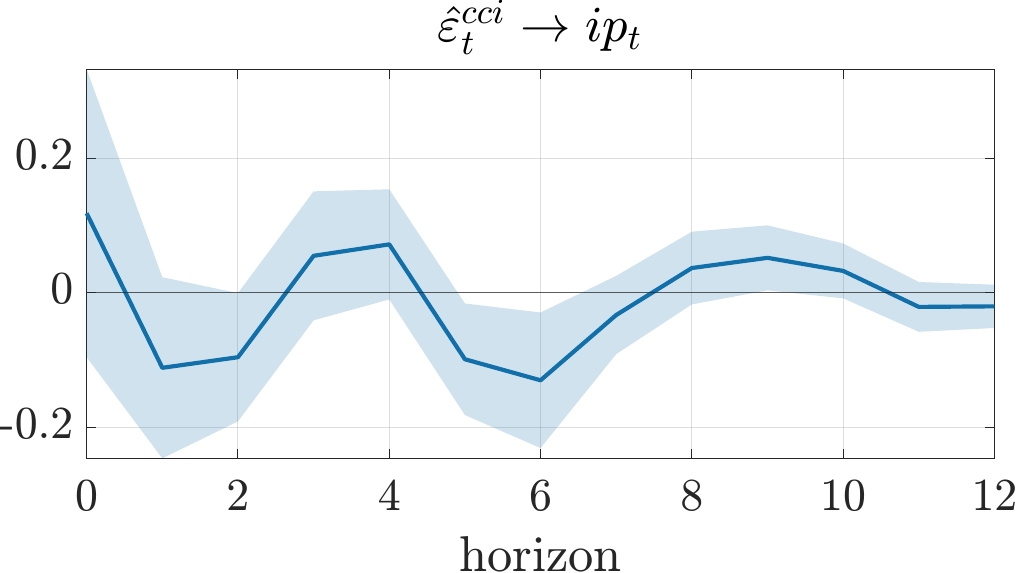}} & {\includegraphics[width=5cm,keepaspectratio]{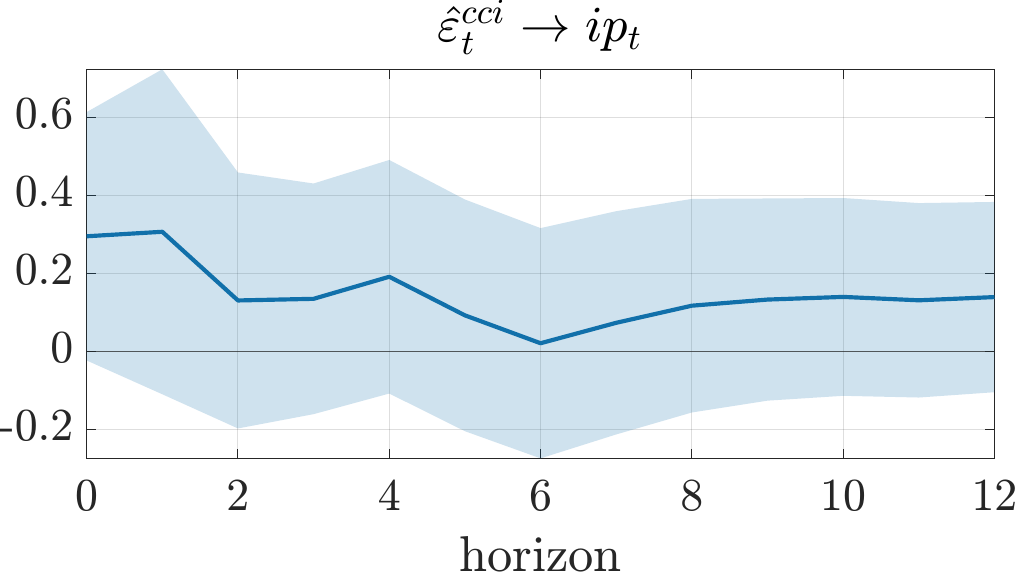}}   \\
{\includegraphics[width=5cm,keepaspectratio]{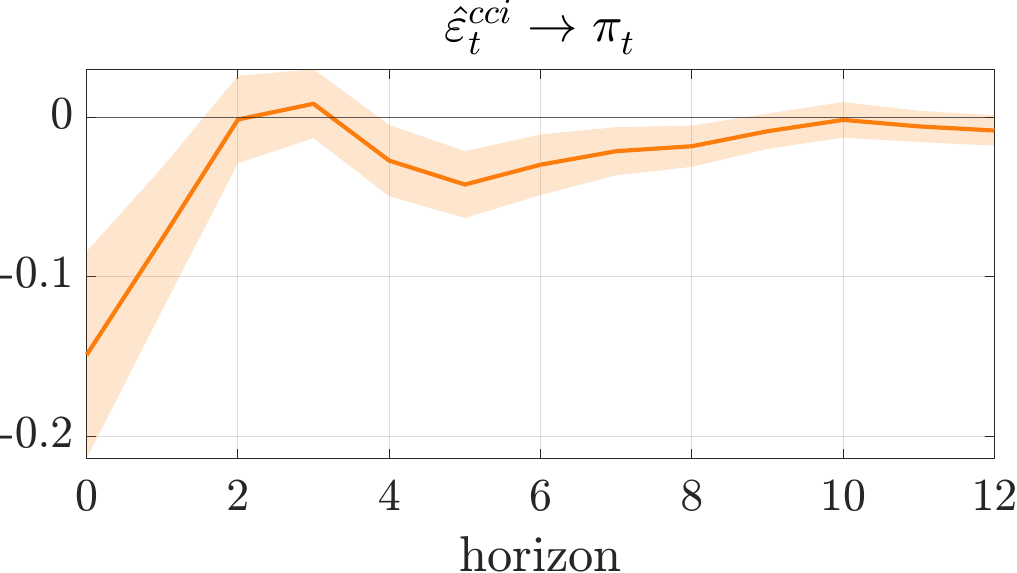}} & {\includegraphics[width=5cm,keepaspectratio]{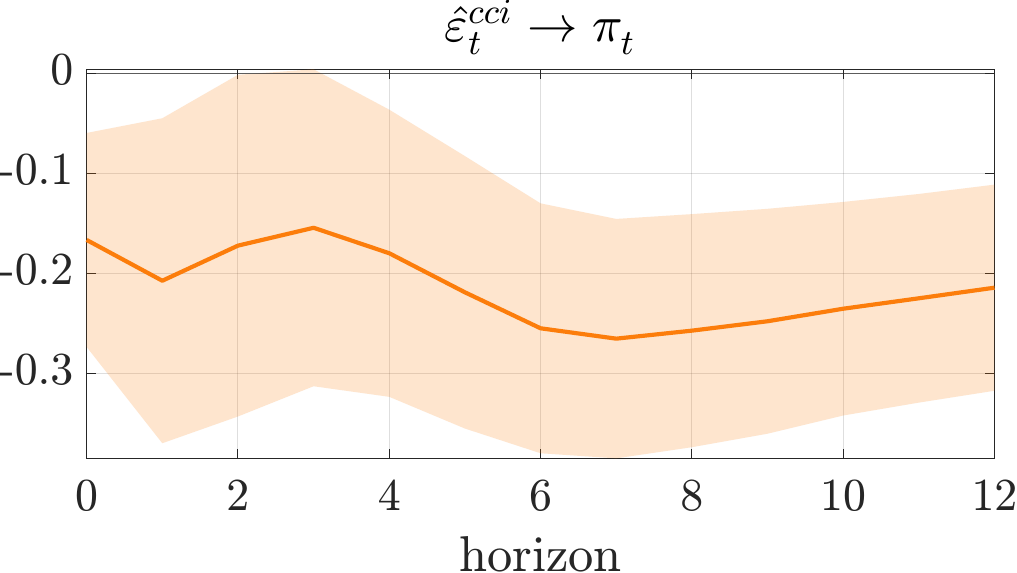}}   \\
{\includegraphics[width=5cm,keepaspectratio]{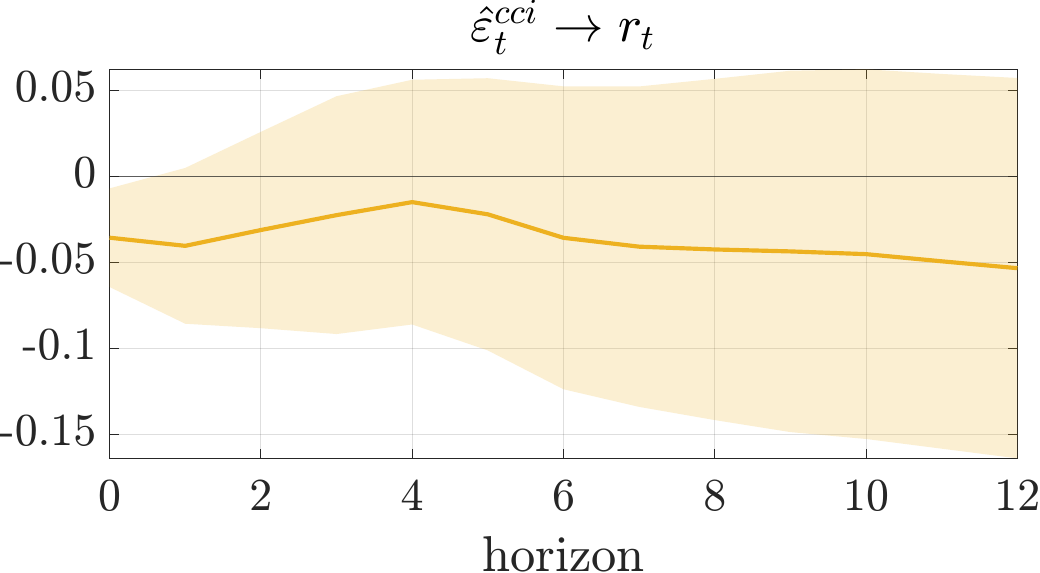}} & {\includegraphics[width=5cm,keepaspectratio]{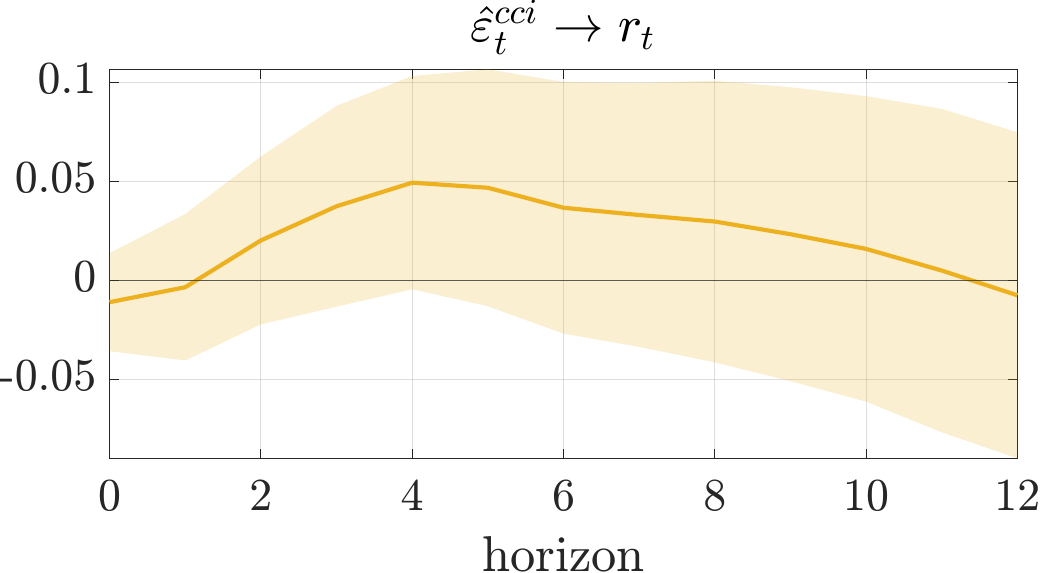}}   \\
{\includegraphics[width=5cm,keepaspectratio]{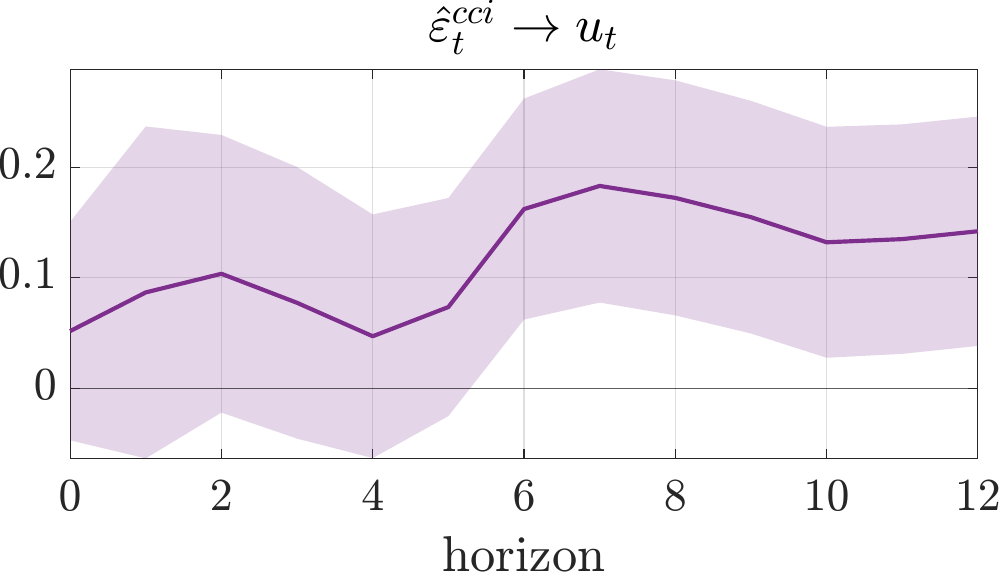}} & {\includegraphics[width=5cm,keepaspectratio]{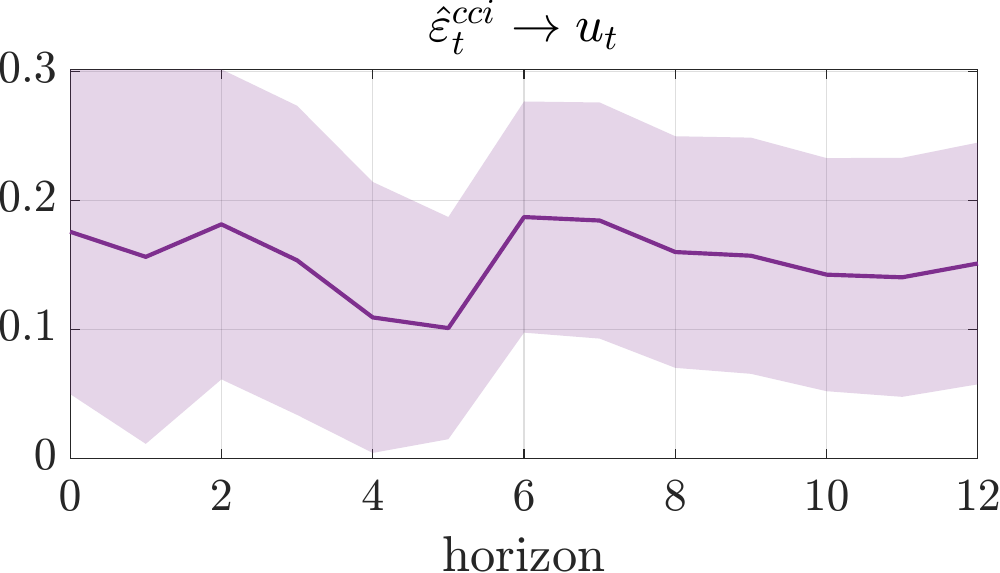}}   \\
\bottomrule
\end{tabular}
\end{center}
\caption{Estimated dynamic effects with 68\% MBB (green shaded area)
confidence intervals at a 12-months horizon, estimated on the sample 2004:M1--2023:M11, of a standard deviation shock on climate concern measure where $Y_t = \left(cci_t,ip_t,\pi_t^{c},r_t,u_t \right)$. On the right the data are year-on-year growth rate.}
\label{fig:irfs_2}
\end{figure}

\begin{figure}[ht!]
\begin{center}
\begin{tabular}{cc}
\toprule
{\includegraphics[width=5cm,keepaspectratio]{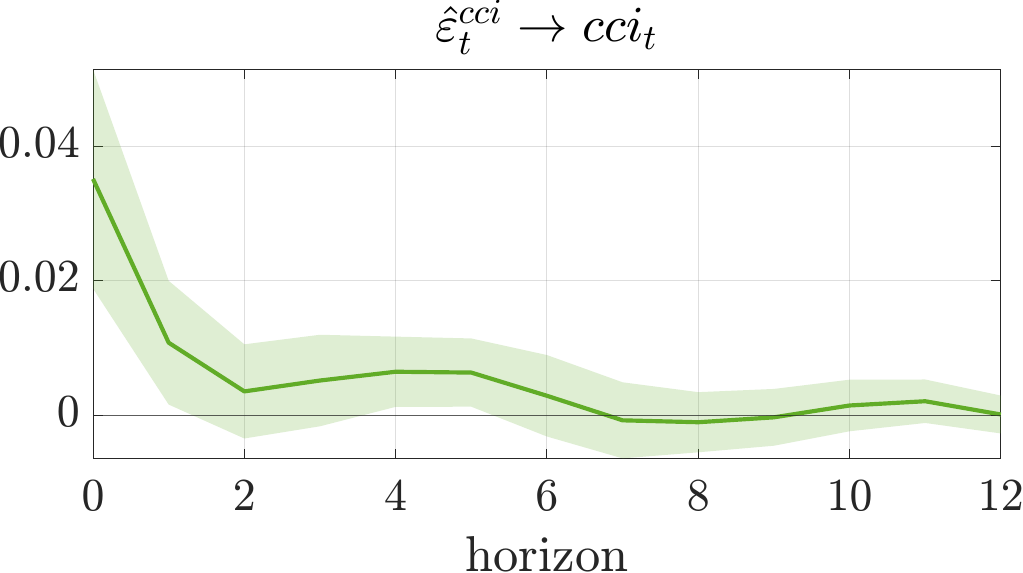}} & {\includegraphics[width=5cm,keepaspectratio]{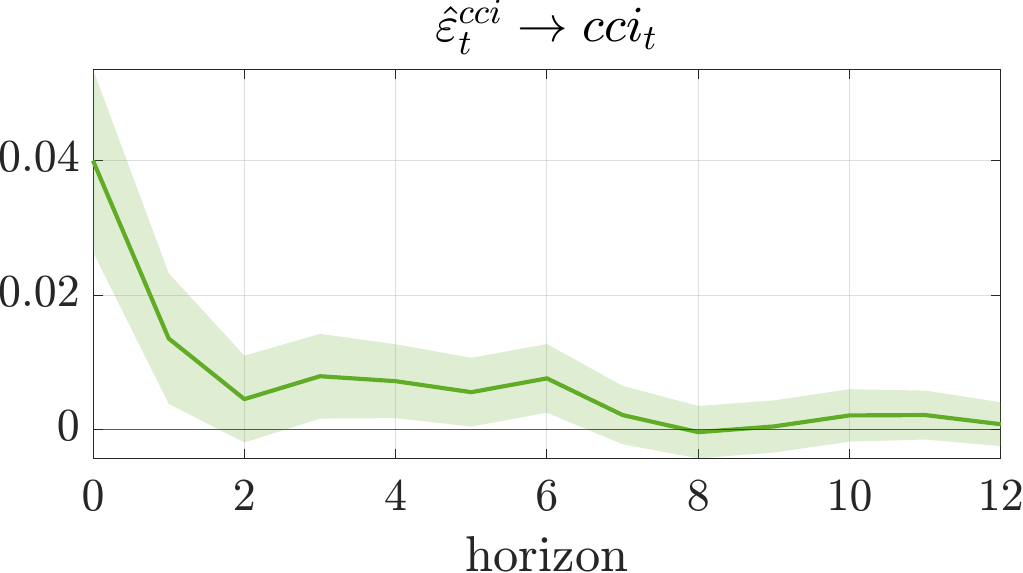}}   \\
{\includegraphics[width=5cm,keepaspectratio]{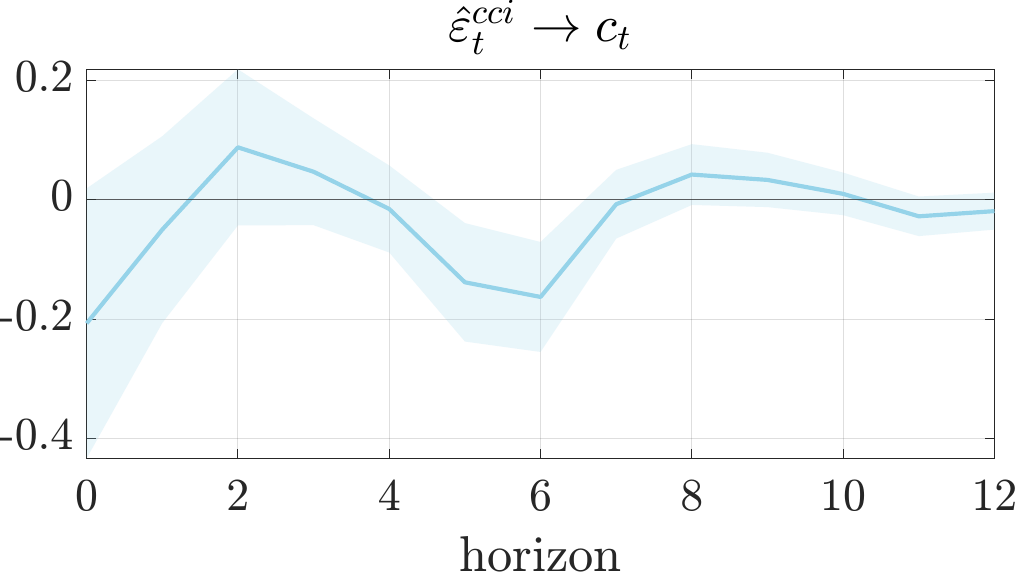}} & {\includegraphics[width=5cm,keepaspectratio]{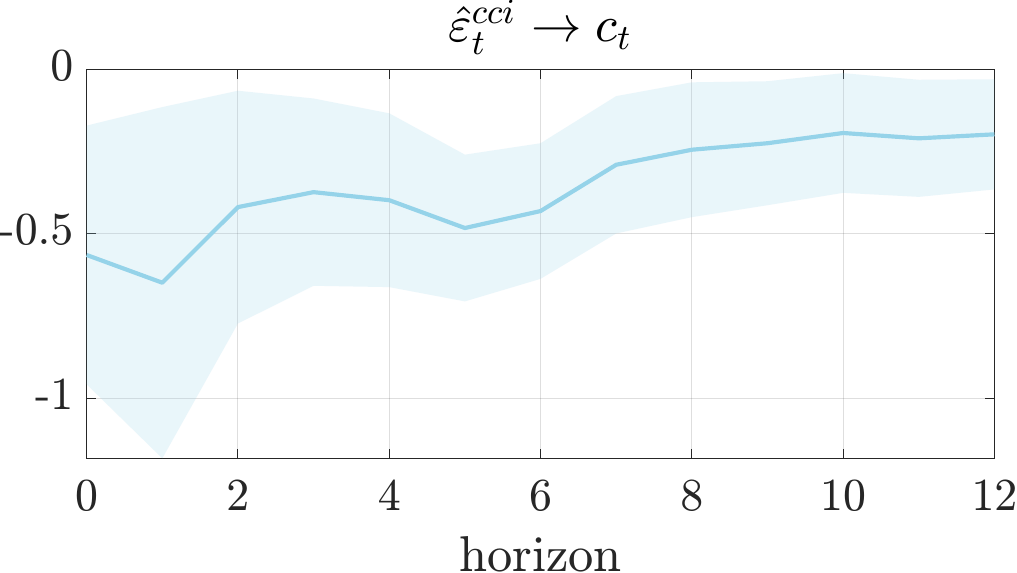}}   \\
{\includegraphics[width=5cm,keepaspectratio]{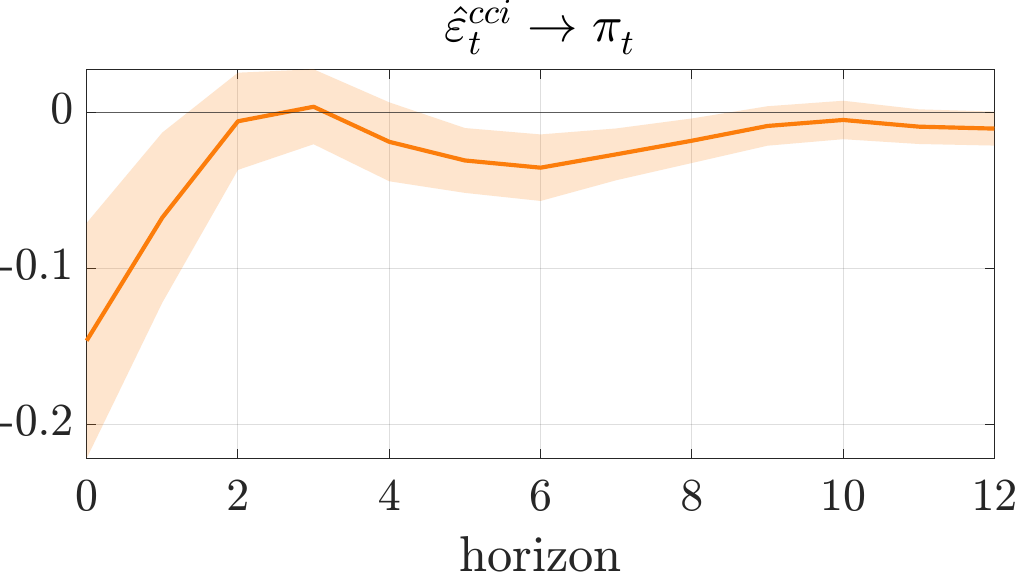}} & {\includegraphics[width=5cm,keepaspectratio]{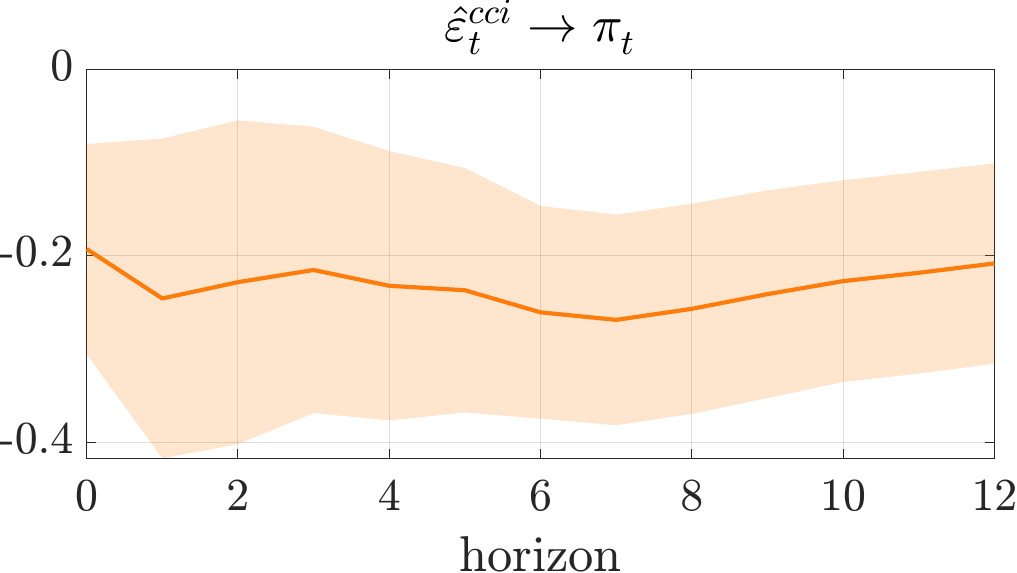}}   \\
{\includegraphics[width=5cm,keepaspectratio]{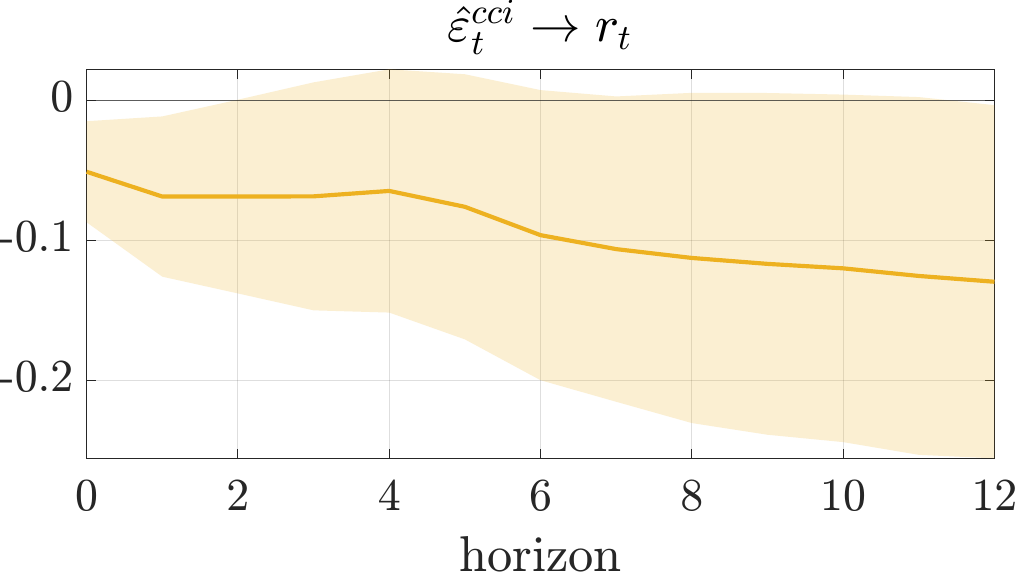}} & {\includegraphics[width=5cm,keepaspectratio]{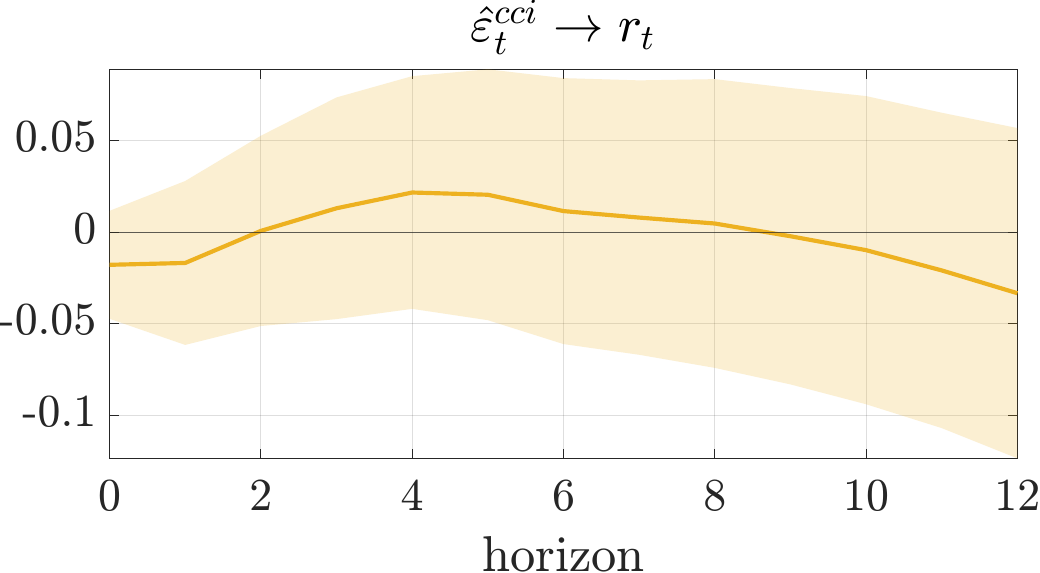}}   \\
{\includegraphics[width=5cm,keepaspectratio]{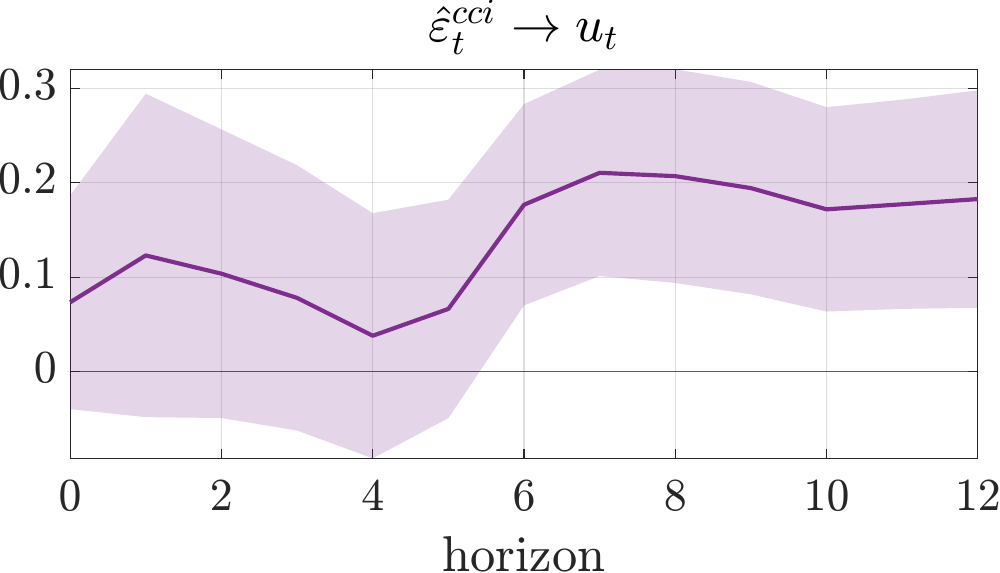}} & {\includegraphics[width=5cm,keepaspectratio]{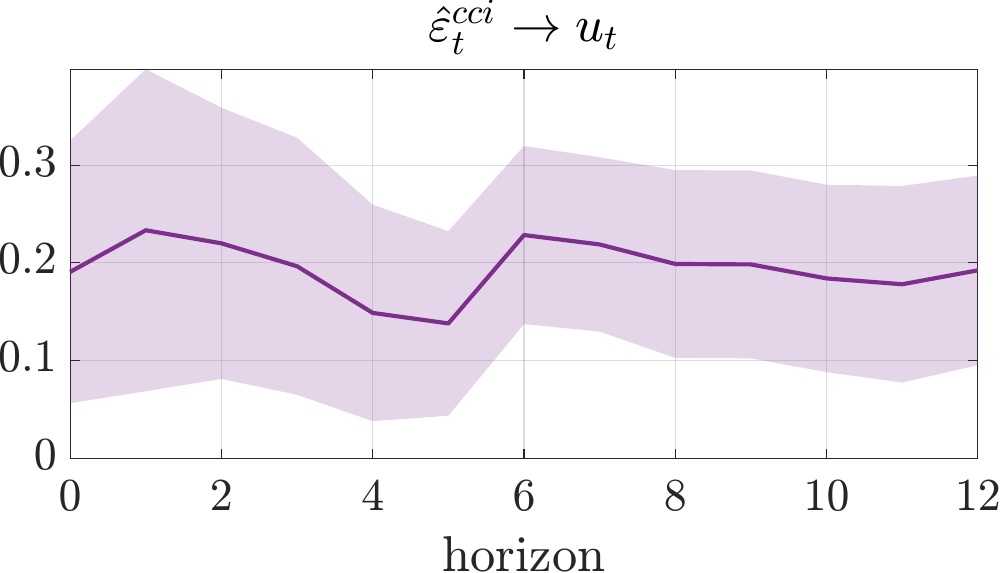}}   \\
\bottomrule
\end{tabular}
\end{center}
\caption{Estimated dynamic effects with 68\% MBB (green shaded area)
confidence intervals at a 12-months horizon, estimated on the sample 2004:M1--2023:M11, of a standard deviation shock on climate concern measure where $Y_t = \left(cci_t,cons_t,\pi_t^{},r_t,u_t \right)$. On the right the data are year-on-year growth rate.}
\label{fig:irfs_3}
\end{figure}

\begin{figure}[ht!]
\begin{center}
\begin{tabular}{cc}
\toprule
{\includegraphics[width=5cm,keepaspectratio]{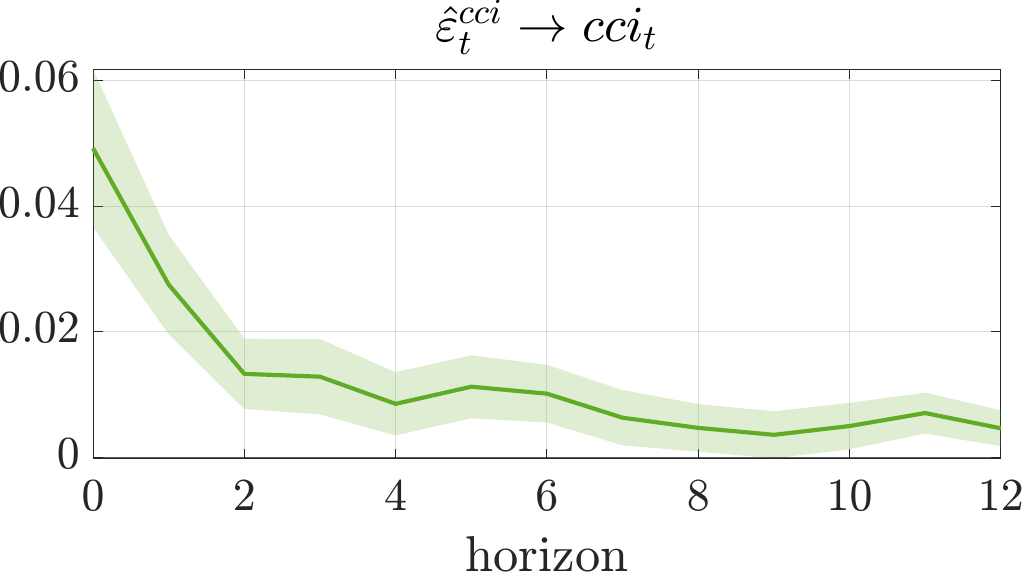}} & {\includegraphics[width=5cm,keepaspectratio]{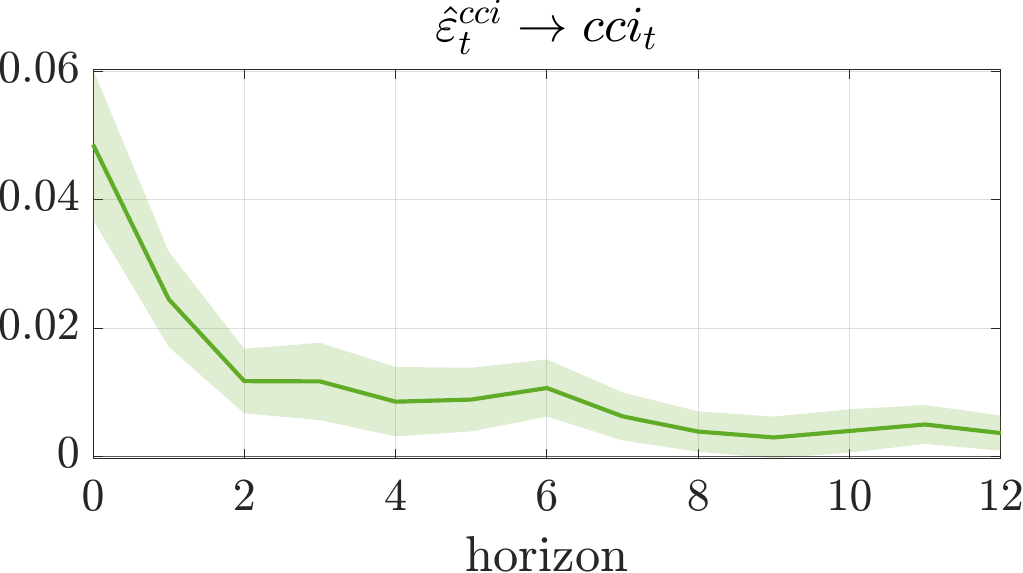}}   \\
{\includegraphics[width=5cm,keepaspectratio]{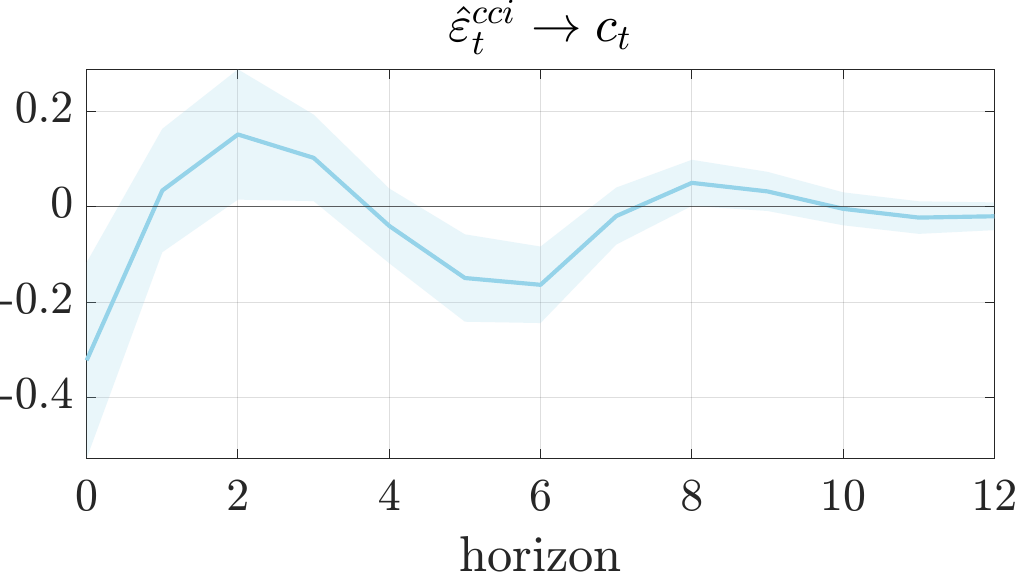}} & {\includegraphics[width=5cm,keepaspectratio]{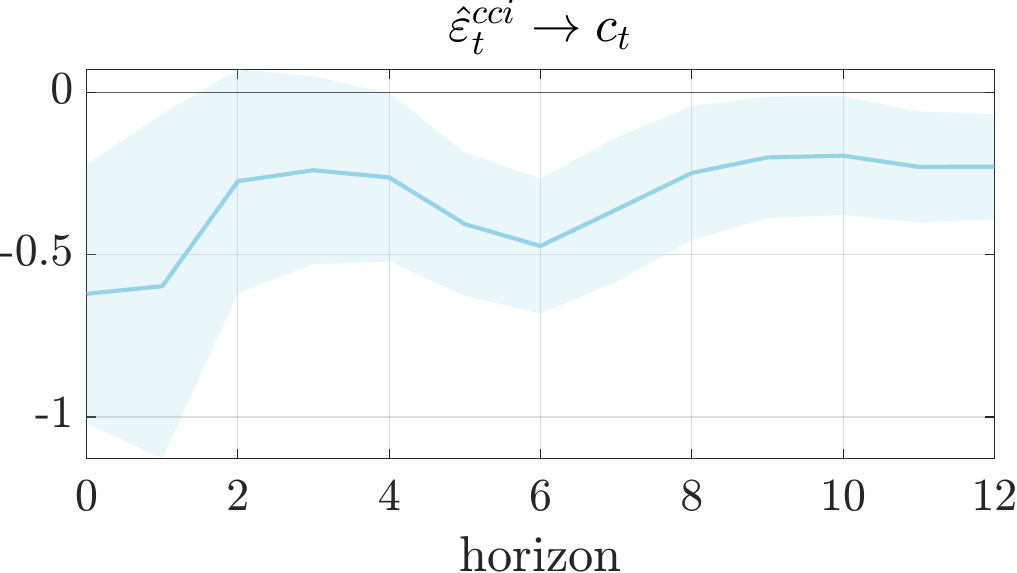}}   \\
{\includegraphics[width=5cm,keepaspectratio]{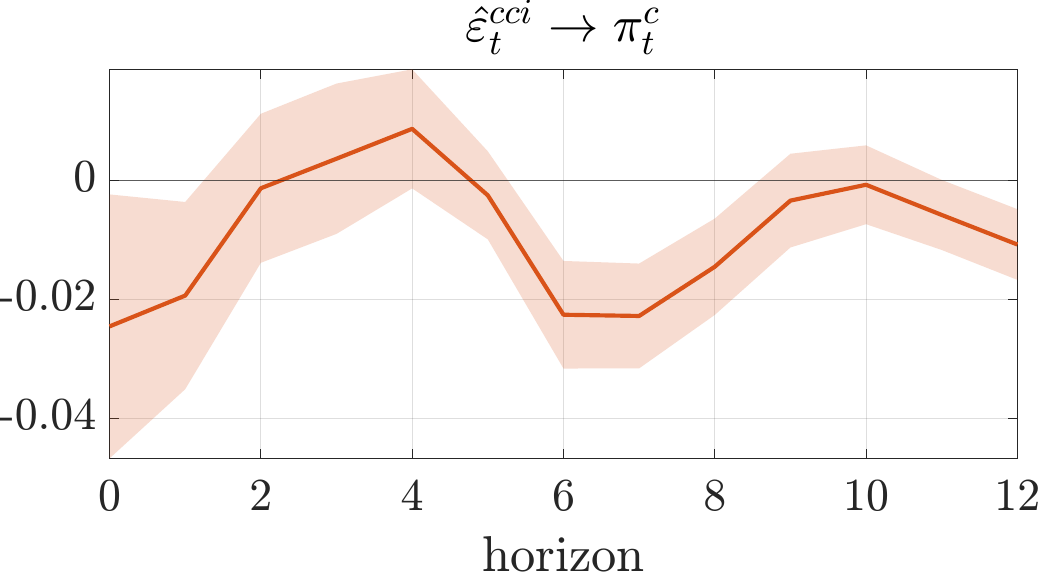}} & {\includegraphics[width=5cm,keepaspectratio]{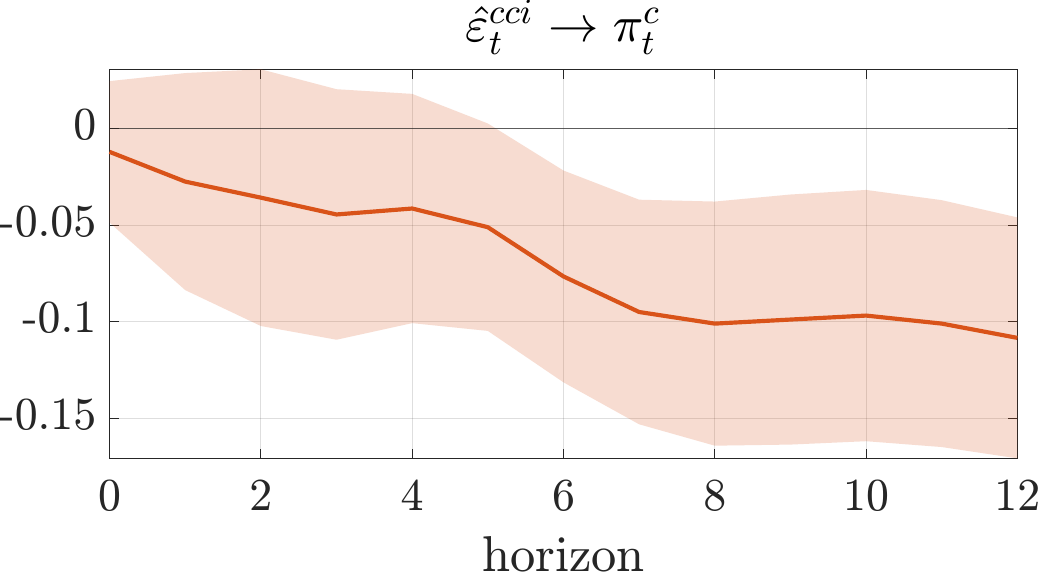}}   \\
{\includegraphics[width=5cm,keepaspectratio]{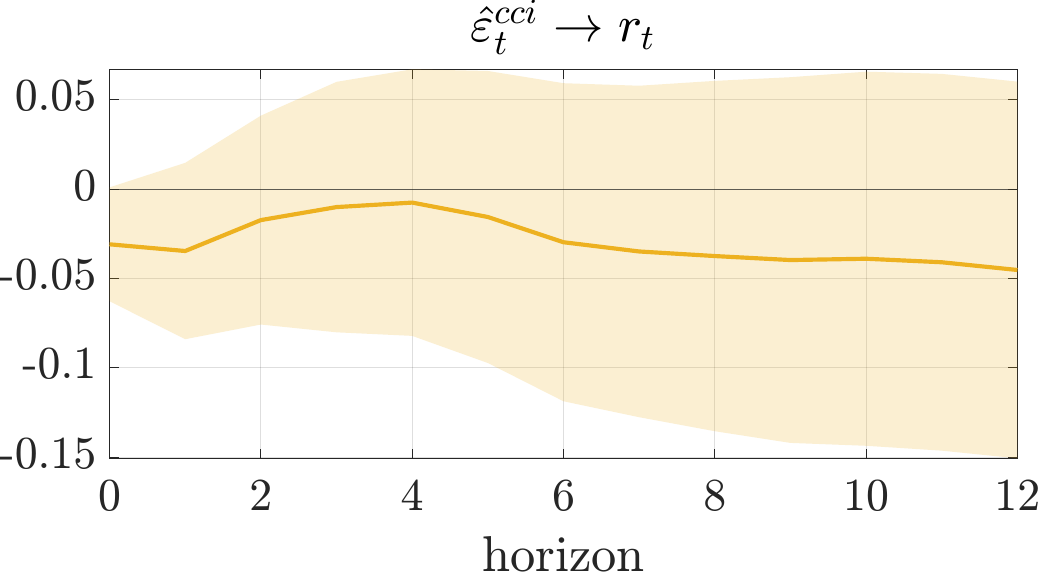}} & {\includegraphics[width=5cm,keepaspectratio]{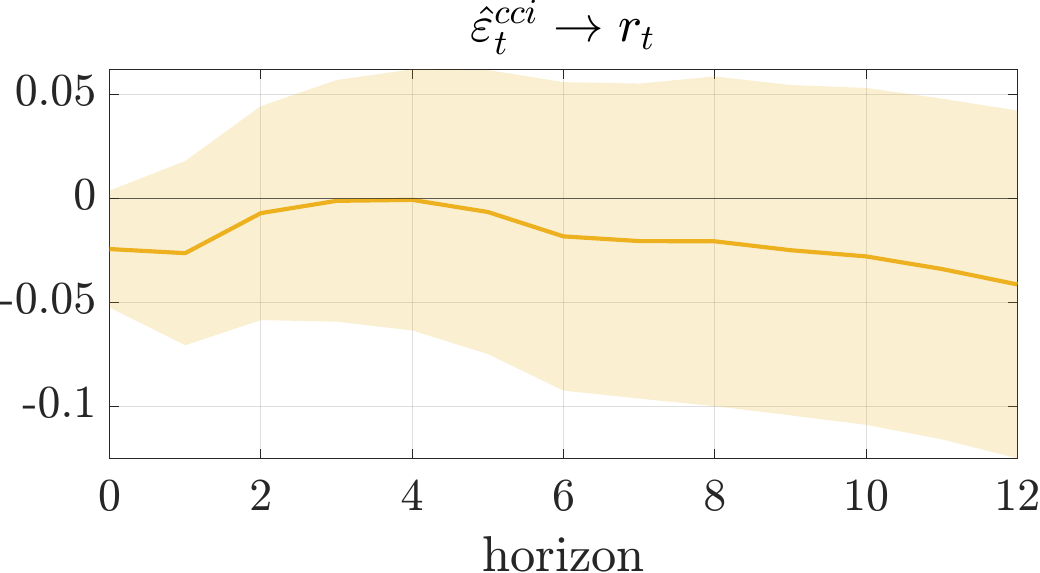}}   \\
{\includegraphics[width=5cm,keepaspectratio]{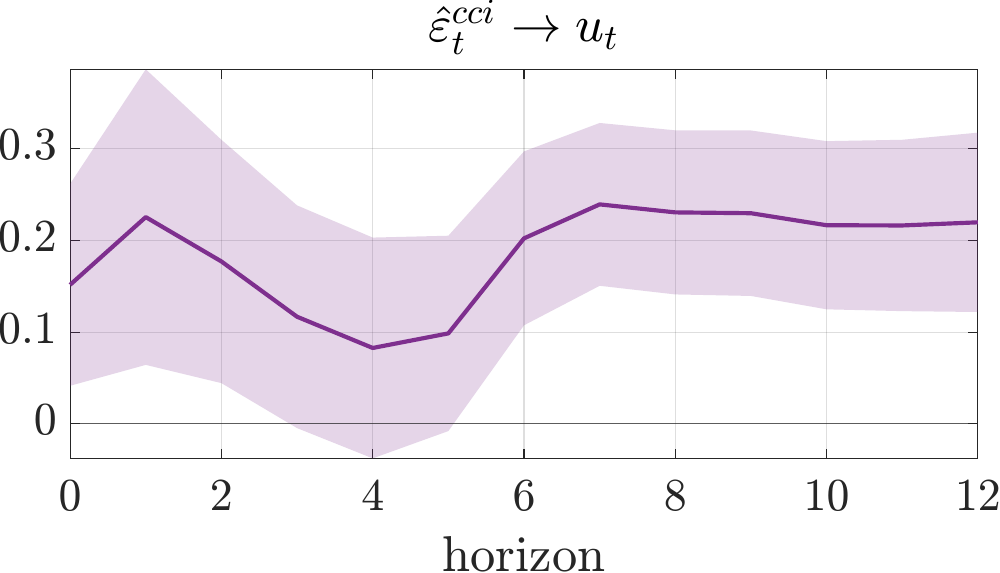}} & {\includegraphics[width=5cm,keepaspectratio]{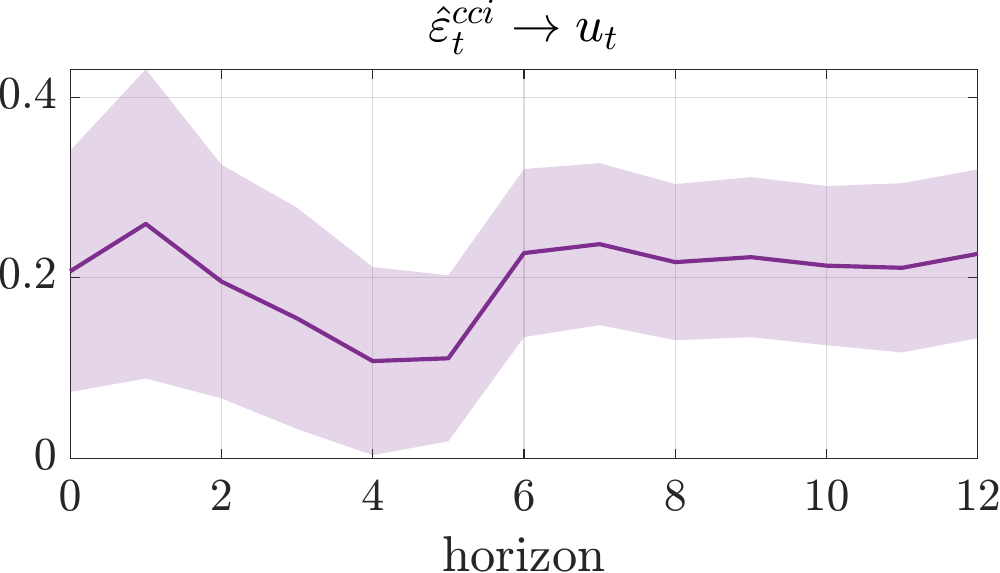}}   \\
\bottomrule
\end{tabular}
\end{center}
\caption{Estimated dynamic effects with 68\% MBB (green shaded area)
confidence intervals at a 12-months horizon, estimated on the sample 2004:M1--2023:M11, of a standard deviation shock on climate concern measure where $Y_t = \left(cci_t,cons_t,\pi_t^{c},r_t,u_t \right)$. On the right the data are year-on-year growth rate.}
\label{fig:irfs_4}
\end{figure}

\begin{figure}[ht!]
\begin{center}
\begin{tabular}{cc}
\toprule
{\includegraphics[width=5cm,keepaspectratio]{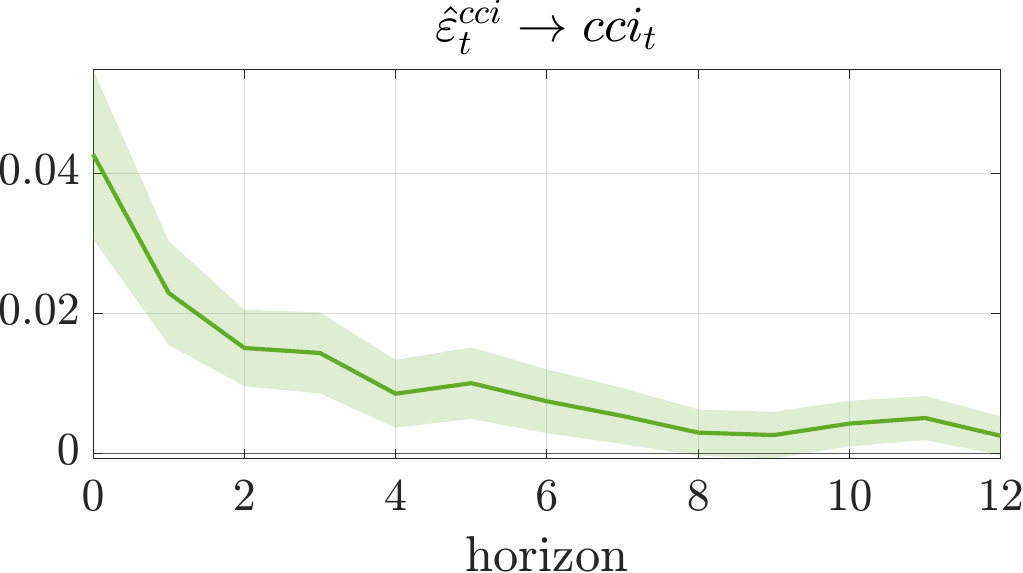}} & {\includegraphics[width=5cm,keepaspectratio]{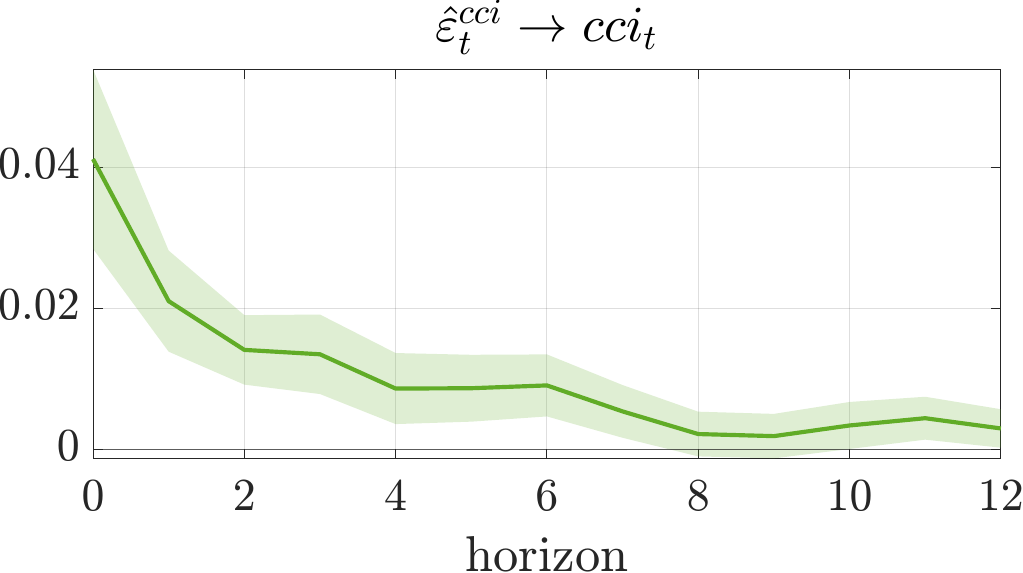}}   \\
{\includegraphics[width=5cm,keepaspectratio]{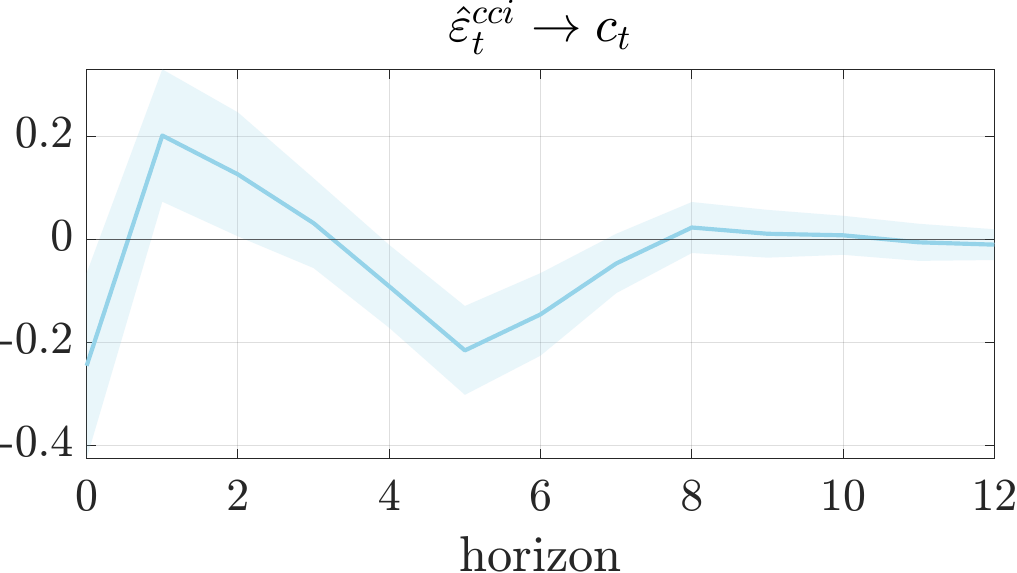}} & {\includegraphics[width=5cm,keepaspectratio]{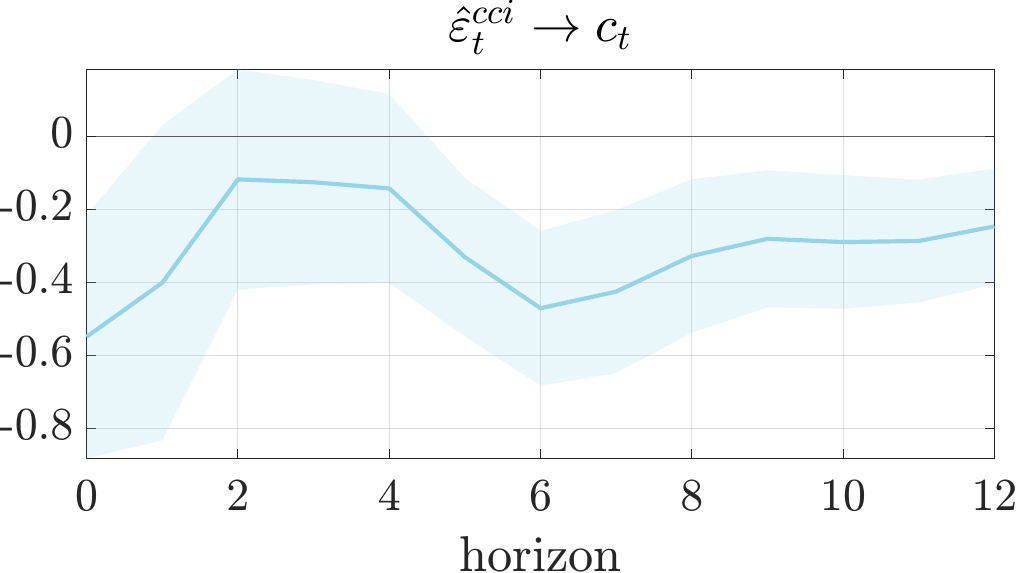}}   \\
{\includegraphics[width=5cm,keepaspectratio]{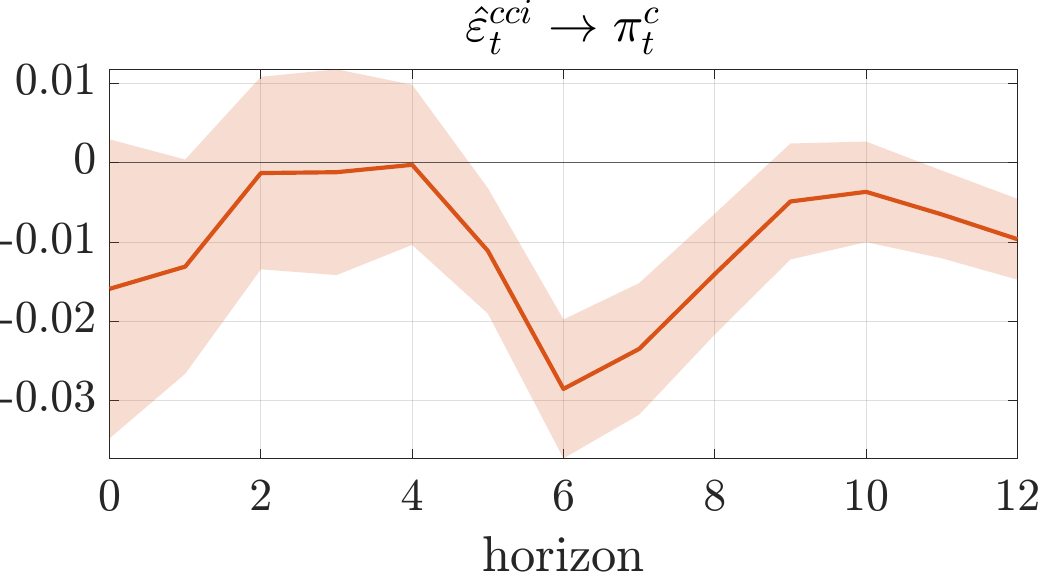}} & {\includegraphics[width=5cm,keepaspectratio]{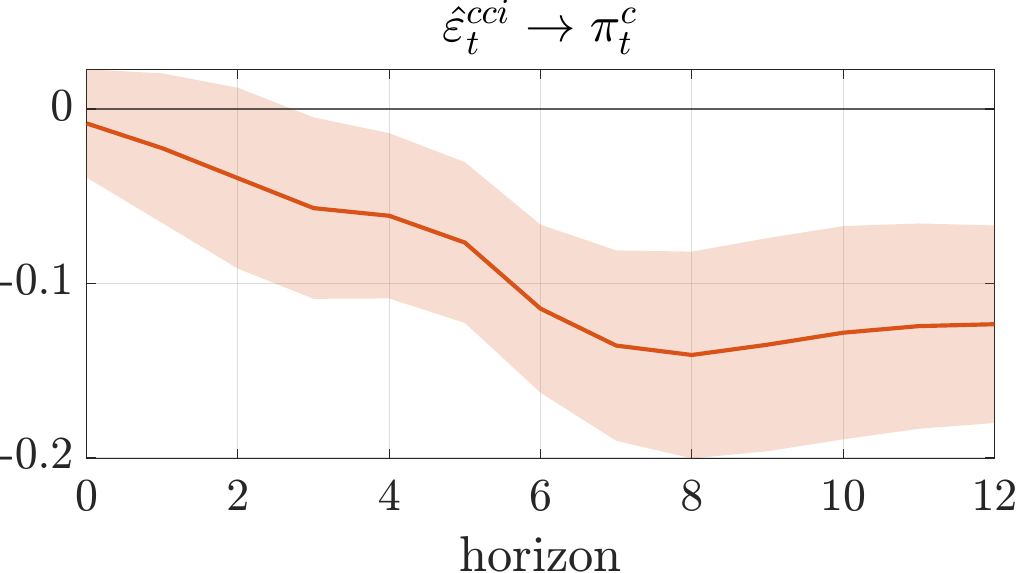}}   \\
{\includegraphics[width=5cm,keepaspectratio]{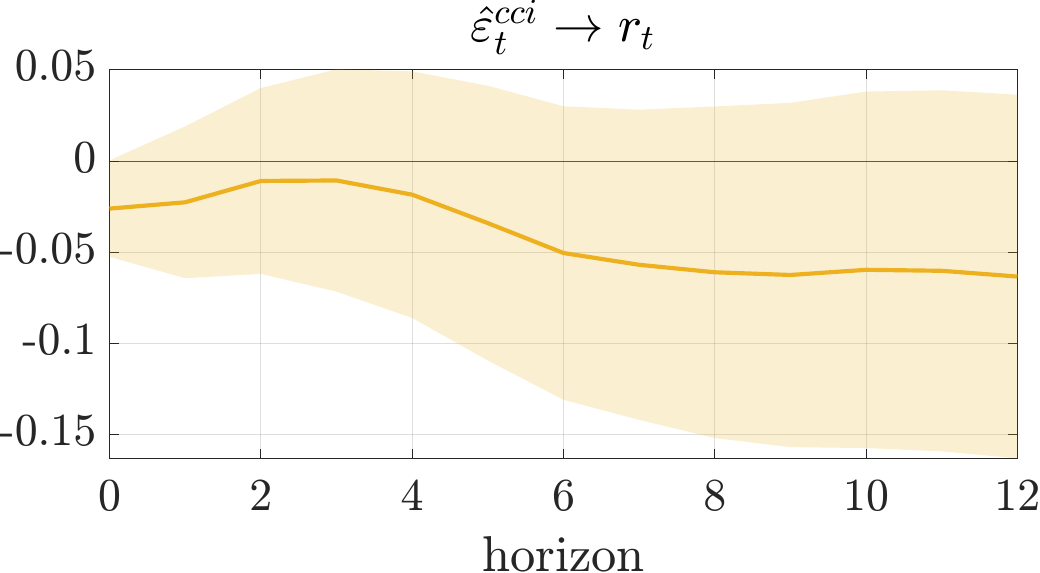}} & {\includegraphics[width=5cm,keepaspectratio]{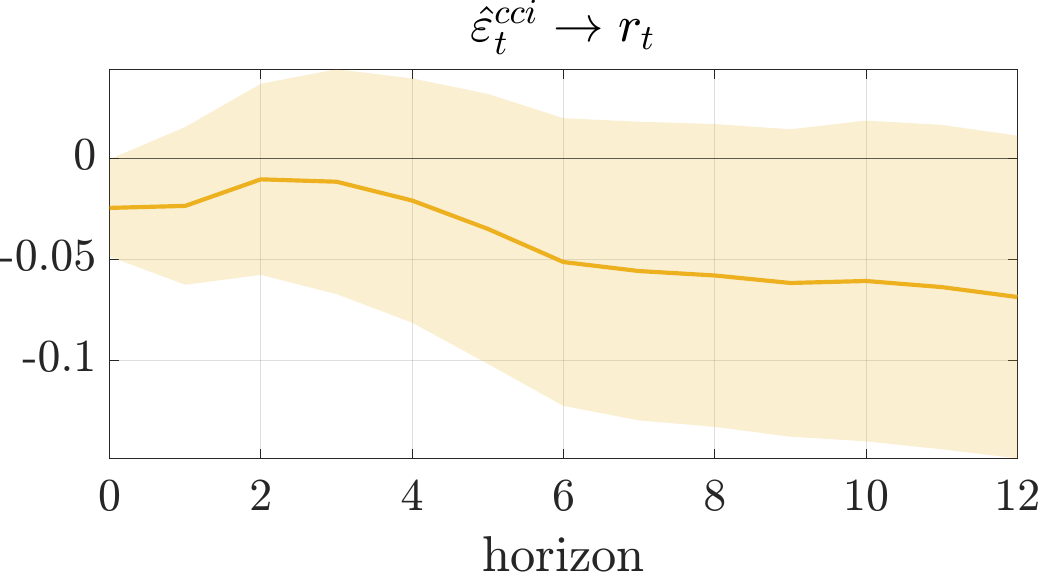}}   \\
{\includegraphics[width=5cm,keepaspectratio]{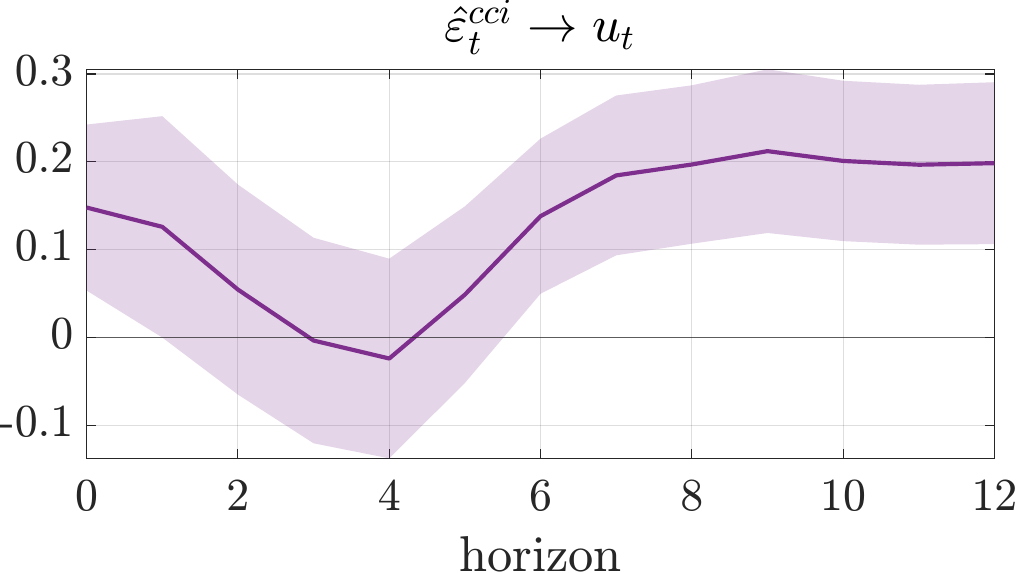}} & {\includegraphics[width=5cm,keepaspectratio]{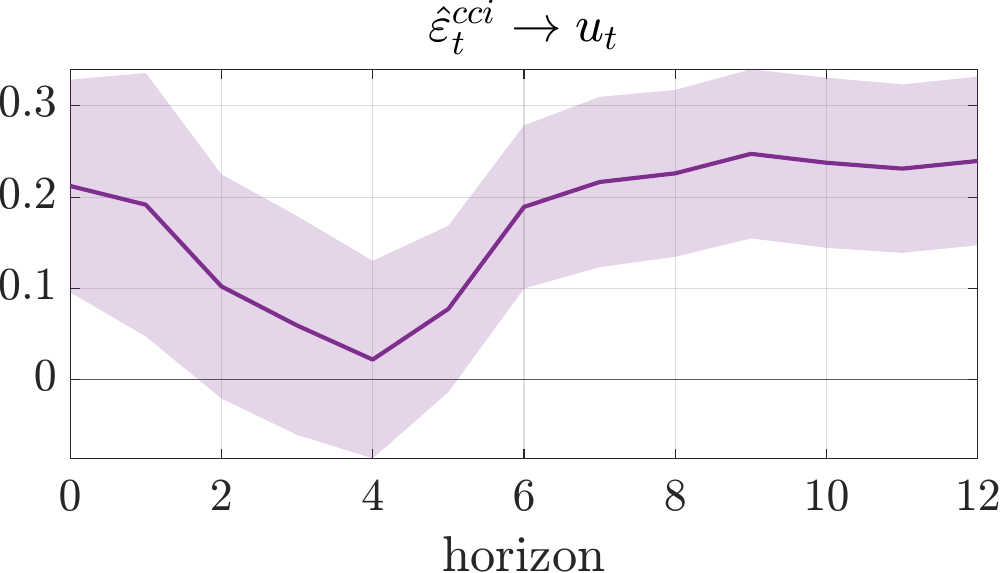}}   \\
{\includegraphics[width=5cm,keepaspectratio]{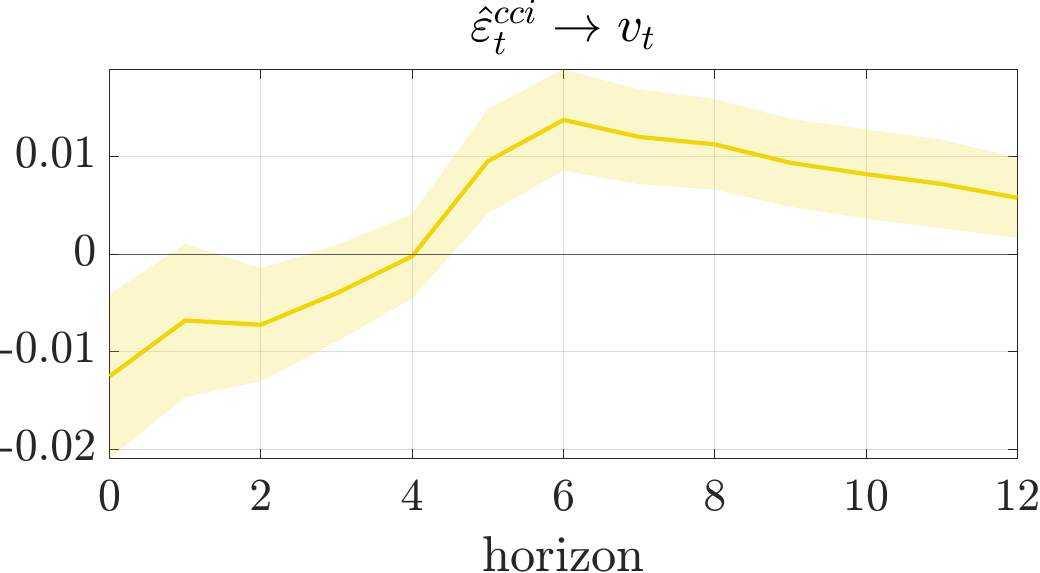}} & {\includegraphics[width=5cm,keepaspectratio]{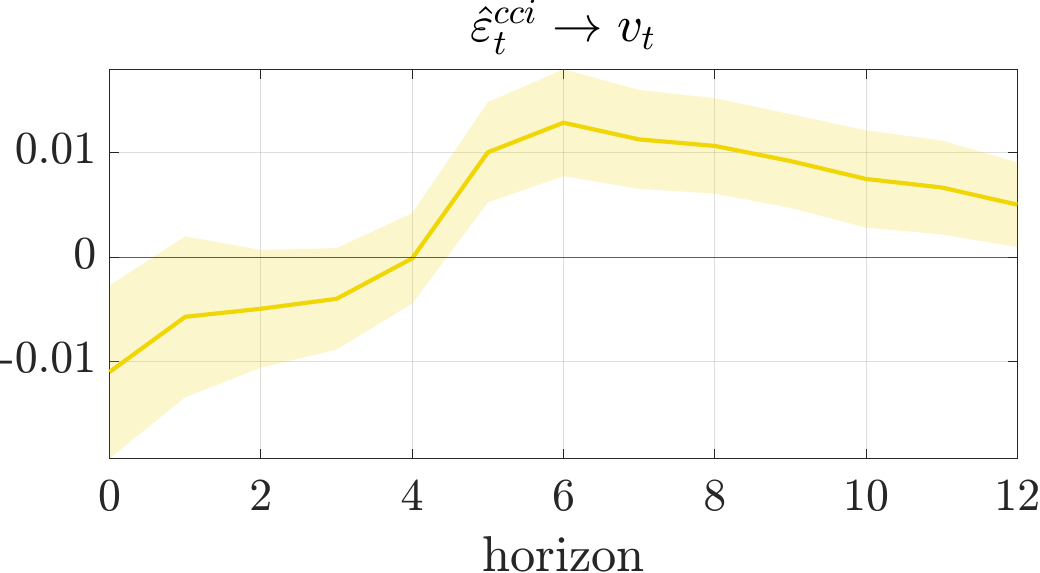}}   \\
\bottomrule
\end{tabular}
\end{center}
\caption{Estimated dynamic effects with 68\% MBB (green shaded area)
confidence intervals at a 12-months horizon, estimated on the sample 2004:M1--2023:M11, of a standard deviation shock on climate concern measure where $Y_t = \left(cci_t,cons_t,\pi_t^{c},r_t,u_t,v_t \right)$. On the right the data are year-on-year growth rate.}
\label{fig:irfs_5}
\end{figure}

Figures (\ref{fig:irfs_2})-(\ref{fig:irfs_5}) report the estimated impulse response functions (IRFs) for a one-time, one-standard-deviation shock to the CCI on a twelve months horizon. We shall remark here that, while our sVAR methodology captures dynamic interactions and spillover effects across the variables of interest, it does not allow us to single out the key underlying mechanisms, shaped by different factors such as households' preferences for savings against non-insurable risk, firms' investment strategies in the face of climate-related expected damages, the relative level of economic development and diversification of the US economy vis-\`a-vis other countries, and US's fiscal and institutional capacity to adapt to and mitigate the consequences of climate change. Nonetheless, our results offer novel evidence on persistent, non-trivial effects of climate risk perceptions on economic activity, price volatility and market uncertainty, and can thus be useful for both disciplining the calibration of theoretical frameworks aimed at evaluating the welfare properties of climate targets, and informing policy discussion about public intervention in private adaptation markets.
\\

\noindent{\textbf{Industrial production, consumption growth and unemployment}}

Irrespective of whether monthly or year-on-year growth rates are considered, the CCI shock bears no statistically significant effects on IP growth, at any time horizon within the considered time span --  see Figure (\ref{fig:irfs_2}). Much in line with \cite{kim2025severe}, this evidence might reflect the fact that global climate change (in particular, increases in temperature) generates wide-ranging effects on agricultural output (e.g. crop yields), which is not part of the industrial production series. Looking at measures of private consumption spending (proxied by FRED's PCE series), which amounts to roughly 50 percent of GDP on average in our sample, reveals a rather different picture: unexpected variation in climate concerns adversely impacts aggregate consumption growth at the beginning of the sample, while inducing non-monotonic adjustment paths towards its long-run value. Most notably, the CCI shock also engenders a non-trivial and persistent rise in the unemployment rate, which exhibits rather slow reversion to its initial level -- see Figures (\ref{fig:irfs_2})-(\ref{fig:irfs_4}). Clearly, climate hazards associated with environmental degradation adversely affect jobs depending on ecosystem services (e.g. farming,	fishing	and	forestry jobs), while the supply of safe and healthy working conditions critically hinges on the preservation of environmental stability. The empirical patterns we uncover coherently suggest that environment-related concerns and uncertainty associated with unfavorable scenarios hinder labor productivity growth and induce higher precautionary savings on the part of households who cannot insure against climate risk. 
\\

\noindent{\textbf{Inflation vs. core inflation}}

Our estimates indicate that the CCI shock has a statistically significant, non-monotonic effect on headline inflation over the entire time span -- see Figures (\ref{fig:irfs_2}) and (\ref{fig:irfs_3}). Following \cite{kim2025severe}, we then look at the joint dynamics of the CCI series vis-\`a-vis core inflation, and find that the response of the latter on impact and over nearly any horizon vanishes, when monthly growth rate for consumption and prices are considered -- see left panels in Figure (\ref{fig:irfs_4}), while it declines below its initial level with some lag and exhibits some persistence in the year-on-year growth specification -- right panels in Figure (\ref{fig:irfs_4}). While suggesting that the response of the CPI inflation in the shorter term is likely driven by variation in energy and food prices, both reacting to extreme temperature shocks (our proxy for unexpected movements in the CCI), the disinflationary effect reported in the right panel of Figure (\ref{fig:irfs_4}) might be due to transmission channels that strongly depend on country-specific factors, on both the market and public sector side, that our empirical model does not explicitly encompass (e.g. income inequality, see \cite{cevik2023eye}).
\\

\noindent{\textbf{Interest rate}}

For all of our specifications, the CCI shock produces no statistically significant effect on the short-term interest rate, in line with the view that monetary policy does not directly or indirectly respond to varying collective concerns about climate change, insofar as they generate mild to negligible inflationary pressure.
\\

\noindent{\textbf{Stock market volatility}}

By shaping investor sentiment, extreme shock events such as severe weather phenomena and climate change-related ones (e.g. microbial growth), on the one hand, and uncertainty surrounding the adoption of effective mitigation policies and incentive measures addressing different economic sectors, on the other, have been empirically shown to trigger non-negligible fluctuations in financial asset prices in advanced market economies, see e.g. \cite{liang2024impacts}. Heightened stock market volatility, in turn, can entail significant macroeconomic implications by distorting the allocation of financial capital, impairing the stable operation of the financial system and possibly hampering economic development and growth \citep{campiglio2025climate}. As financial stability may exhibit time-varying patterns of association with price stability \citep{blot2015assessing}, and in view of our previous discussion about the inflationary effects of climate-related concerns, uncovering the dynamic response of stock market volatility to varying climate awareness would provide additional evidence-based guidance for the policy debate about financial (de)regulation and the design of macro-prudential measures.

To gain insight into this issue, we augment the VAR with a standard measure of expected volatility of the US stock market, i.e. the  Chicago Board Options Exchange Volatility Index (VIX). Based on the option prices of the S\&P 500 Index, the VIX is explicitly designed to gauge investors' collective expectations of the variability of future market movements over a 30-day time window. Since routinely used as an indicator of market fear and stress, we seek to explore whether CCI shocks may play a role in fueling stock market uncertainty via the investor sentiment channel. As Figure (\ref{fig:irfs_5}) suggests, and in line with the existing literature on this specific issue, e.g. \cite{li2024study}, this appears to be case: albeit with some delay, possibly due to slow factoring of perceived climate-related risks into asset prices through the investor sentiment channel, the response of expected market volatility to public perceptions of climate change is positive and hump-shaped, reaching a peak (about $1\%$) at the sixth month after the shock and then slowly reverting back to the mean in the following periods. 

\clearpage
\newpage

\clearpage
\newpage

\clearpage
\newpage

\clearpage
\newpage

\section{Conclusion}\label{CONCL}

As long as public perceptions of climate change are likely to trigger non-negligible behavioral responses that in tun affect both the economic landscape and the environment, understanding the nature and the dynamics of these perceptions can help us qualify the different mechanisms at play, to tailor communication strategies that allow people to make informed decisions, but also to devise and target environmental and other kinds of socio-economic policies to the most vulnerable groups in the population. 
 
Our web search-based indicator quantifies the intensity and dynamics of public concerns over time. Using data from the United States over the 2004–2024 period, it offers novel insights into collective and widespread shifts in perceived climate-related risks, that appears not to be incorporated in competing alternatives based on newspapers articles or messagging platforms. Moreover, estimates from a proxy-SVAR model of the US macroeconomy indicate that exogenous variation in the CCI leads to significant declines in employment and private consumption, while also fueling stock market uncertainty as proxied by the VIX index. These results suggest that heightened concerns for climate change can trigger behavioral adaptation with tangible macroeconomic consequences, even in the absence of direct physical risk or efforts to mitigate it. Since climate policy efficacy, whether market-based or not, strongly hinges on the general audience's understanding of the intervention and willingness to support, our analysis also calls for carefully designing communication strategies and risk management tools able to mitigate climate-related uncertainty and discipline the formation of collective concerns about climate change.

%\section*{Acknowledgments}
%We wish to thank Matteo Ciccarelli, Elena Paglialunga and participants at several workshops, conferences and department seminars where early versions of our paper have been presented. Financial support from MUR (PRIN grant 2022H2STF2) is gratefully acknowledged.

\newpage

\bibliographystyle{chicago}
\bibliography{References}

\newpage
\appendix

 % appendix (put online supplement in a separate file)
% put S in front of counters for online supplement
\renewcommand{\theequation}{A.\arabic{equation}}
\renewcommand{\thesection}{A.\arabic{section}}
\renewcommand{\thepage}{A.\arabic{page}}
\renewcommand{\thetable}{A.\arabic{table}}
\renewcommand{\thefigure}{A.\arabic{figure}}

\setcounter{equation}{0}
\setcounter{page}{1}
\setcounter{table}{0}
\setcounter{equation}{0}
\setcounter{section}{0}
\setcounter{figure}{0}

\section{Appendix}\label{APP}
\subsection{Dictionary of the CCI index}\label{sez:APPdic}

\begin{table}[!htp] 
\caption{Dictionary: categories 1-4}\label{tab:dic1} 
\centering
\tiny
\begin{tabular}{l}
\toprule
\multirow{1}*{}{\textbf{1. Natural disasters that stimulate fear (negative)}} \\
Extreme weather -earth -day -research -study -cold -heat -very -hot -cold -near -me -today \\ 
Deepwater horizon -movie -deep \\
Floods  -research -essay -adventures -movie \\
Storm snow + snowstorm -picture -painting -research -synonym -movie -film -update -weather -game \\ 
Tornadoes -movie -film -near -me -warning -today -update -weather -game \\ 
California fires -fake -reddit -pg\&e -costs -insurance \\ 
Blizzard -menu -twitter -diary -overwatch -essay \\
Hurricanes -football -team -lyrics -hockey -Carolina -Jerseys -tickets -garden \\ 
\midrule
\multirow{2}*{}{\textbf{2. Global warming one type of concern (negative)}} \\
Sea rise causes + sea rise effects \\
Deforestation -tropical -conference -school -research -study -essay \\
Ice caps melting \\
Global average temperature -up -history -nasa -data -2000 -years -by desertification \\
Help global warming + reduce global warming \\
Extreme cold weather -lyrics -tent -clothing -drink -death -damage -fingers \\
Extreme heat + heatwaves -glass -animals -song -lyrics -disease -stroke -jb -weld -what -to -do \\
El Niño -movie -sanchez -iker -song -guerrero \\
Heathwave + drought \\
AMOC -aston -club \\
\textit{Global warming -research -school -essay} \\
\midrule
\multirow{1}*{}{\textbf{3. Climate change another type of concern (negative)}} \\
Climate refugees \\
Climate crisis -resilience -fee -research -font -variable -advisory -group \\
Climate change impacts + climate change effects \\
Fossil fuels -research -definition -examples -composition -burning -oil \\
Climate change pollution \\
GHG \\
Carbon Dioxide gas -research \\
Greenhouse Effect -earth -day -conference -study -essay -research \\
Petroleum -gas -oasis -oil -marathon -engineering -jelly -research -stock -spot -future -price \\
Environmental impact \\
Yellow vest protest \\
Plastic pollution \\
Fast fashion \\
Greta Thunberg -reseach -age -book -asperger -deasease -swedish -econometrics -worth \\
Earthday -when -where -google -birthday -festival -kids\\
\textit{Climate change -earth -day -where -when} \\
Climategate \\
EPA Vehicle emission \\
Volkswagen dieselgate \\
EPA GHG emissions -calculator \\
\midrule
\multirow{1}*{}{\textbf{4. Climate change reduction /mitigation (positive)}} \\
Climate solutions -earth -day 	\\
Environmental sustainability -research -earth -day 	\\
Reduction climate change -cost -finance 	\\
Energy waste reduction -research -results -poster -drawing 	\\
Renewable energy -earth -day -wallpaper -quotation -community -source -journal -factor -school 	\\
Circular economy -jobs -macarthur -diagram -flow -what -is 	\\
Carbon cycle -earth -day -conference -price -stock 	\\
Energy transition 	\\
Green energy 	\\
Hydrocarbons -composition -study 	\\
Biomass energy 	\\
Reduce carbon emissions -quizlet -poster -jstor -kids -labster -xls 	\\
Thermal insulation -rull -research -resistance -test 	\\
Nitrogen fertilizer 	\\
Green revolution 	\\
Biogas +Biomethane -price 	\\
\textit{Natural gas -propane -down -up -hub -spot -future -shock -price -heating -piedmont -fracking} 	\\
US China agreement climate change 	\\
\bottomrule
\end{tabular}
\end{table}

\newpage
\clearpage

\begin{table}[!htp] 
\caption{Dictionary: categories 5-7}\label{tab:dic2} 
\centering
\tiny
\begin{tabular}{l}
\toprule
\multirow{1}*{}{\textbf{5. New technology, hope (positive)}} \\
Biodiversity -earth -day -conference -essay -research -pronunciation -journal -conservation 	\\
Aquaculture -fish -job -international -research -journal -guide -delight 	\\
Vegetation index 	\\
Cloud seeding 	\\
Solar power -bank -system 	\\
Wind energy -costs -solar 	\\
Ocean energy 	\\
Hydrogen fuel -near -me 	\\
Carbon capture 	\\
Vertical farming 	\\
Electric vehicles -top -best -used -hybrid -tax -bmw -gm -tesla -chevy -nissan 	\\
\midrule
\multirow{1}*{}{\textbf{6. International summits that stimulate interest (neutral)}} \\
Climate action summit 	\\
Paris agreement -policy -countries -nuclear -weapons 	\\
Kyoto Protocol   	\\
Copenhagen Conference 	\\
G8 climate 	\\
Bali climate conference 	\\
Doha conference 	\\
National Climate Assessment 	\\
Marrakech Climate conference 	\\
Montreal climate Conference 	\\
UN Climate Change Conference 	\\
UN climate action summit 	\\
\midrule
\multirow{1}*{}{\textbf{7. Climate/environmental policies that stimulate attention (neutral)}} \\
Climate sustainability -masters -services -companies -certification -fund -job -courses 	\\
Carbon footprint -research 	\\
Decarbonization -engine -stock -companies  -costs 	\\
IPCC 	\\
Carbon credit -up -down -peak -market -price 	\\
Democrats climate agenda 	\\
Senate climate change bill 	\\
Obama announces climate strategy 	\\
Lieberman-Warner Climate Act 	\\
GOP block climate legislation 	\\
GOP climate bill 	\\
GOP push climate agenda 	\\
Obama confirms climate policy 	\\
Pruitt EPA nomination 	\\
Trump climate policy 	\\
Trump withdraws Paris agreement 	\\
Trump Keystone XL 	\\
Senate rejects Green New Deal 	\\
Judge Dakota pipeline 	\\
McCain Climate Stewardship Act 	\\
Bush policy climate 	\\
Energy Bill debate -duke 	\\
Stern against Kyoto protocol 	\\
National Climate Assessment Report 	\\
Obama Keystone XL 	\\
US withdraws Paris agreement 	\\
Trump carbon emissions 	\\
\bottomrule
\end{tabular}
\end{table}

\end{document}